\documentclass [12pt,article,aps,showpacs,floatfix]{revtex4}
\def\be{\begin{eqnarray} &&}
\def\nonu{\nonumber \\ &&}
\def\ee{\end{eqnarray}}
\def\psla{ \slash \! \! \!}

\def\ket#1{\hbox{$\vert #1\rangle$}}   
\def\bra#1{\hbox{$\langle #1\vert$}}   
\def\oneh{{\textstyle {1\over 2}}}
\def\oneq{{\textstyle {1\over 4}}}
\newcommand{\bm}[1] {\mbox{\boldmath{$#1$}}}

\usepackage{graphicx}
\usepackage{epsfig}
\usepackage{amssymb}
\begin{document}

\title{ Pion Generalized Parton Distributions 
with  covariant and Light-front constituent quark
 models}

\author{T. Frederico$^a$, E. Pace$^b$, B. Pasquini$^c$ and
G. Salm\`e$^d$  } \affiliation{ $^a$ Dep. de F\'\i sica, Instituto
Tecnol\'ogico de Aeron\'autica,  12.228-900 S\~ao
Jos\'e dos Campos, S\~ao Paulo, Brazil\\
$^b$ Dipartimento di Fisica, Universit\`a di Roma "Tor Vergata" and
Istituto Nazionale di Fisica Nucleare, Sezione Tor Vergata, Via
della Ricerca Scientifica 1, I-00133  Roma, Italy \\
$^c$ Dipartimento di Fisica Nucleare e Teorica, Universit\`a degli
Studi di Pavia and Istituto Nazionale di Fisica Nucleare, Sezione di
Pavia, Italy \\
$^d$Istituto  Nazionale di Fisica Nucleare, Sezione di Roma, P.le A.
Moro 2, I-00185 Roma, Italy \\
  }

\begin{abstract}
We investigate 
 the model dependence of no-helicity flip    
generalized parton 
distribution of the pion  upon  different approaches for the quark-hadron 
  and  quark-photon vertexes, in the spacelike region.
In order to   obtain information on 
 contributions from both the valence and 
 non-valence regions,  we compare   
 results for   spacelike momentum transfers 
obtained from i) an analytic covariant model with a bare
quark-photon vertex,  ii) a 
Light-front approach with a quark-photon vertex dressed through a microscopic
vector-meson model and iii) a Light-front approach based on the Relativistic Hamiltonian
 Dynamics. 
 Our comparisons lead to infer the same dynamical mechanism, the
 one-gluon-exchange dominance at short distances, as a source of both  
 the electromagnetic form factor at large
momentum transfer 
and  the parton distribution close to the end-points.
The expected  collinear behavior  of the     
generalized parton 
distributions at high momentum transfer, i.e. a maximum for $x\sim 1$, is also illustrated, 
 independently of the different approaches.  
  Finally a comparison with  recent Lattice calculations of the
 gravitational form factors is  presented.
 \end{abstract}

 \maketitle


\section{Introduction}

In recent years a growing interest in the study of the Deeply Virtual Compton 
Scattering (DVCS) has motivated an impressive amount of work aimed at the extraction
of the so-called Generalized Parton Distributions (GPD's) from experimental data
(see, e.g., Refs. 
\cite{Ji98,Rady01,Goeke:2001tz,diehlpr,Belitsky:2005qn,pasquinirev07}  for recent reviews). In principle,
 GPD's allow one to achieve an unprecedented level of detail   on the knowledge
 of hadronic states.

Naturally, the pion GPD should represent a test ground  of  any approach 
that addresses the issue of obtaining
 a detailed description of  hadron structure, and this explains the  wealth 
 of  papers devoted to such
a task (see, e.g., 
\cite{diehlpr,poly,Tiburzi,Mukh02,Theussl,diehl05,Ji,bronio08,bronio08b,
vandyck08}).
In what follows,  we    focus on the
 GPD's that do not depend  upon  the helicities 
of the constituents, namely we   analyze the pion isoscalar and isovector GPD's,
as defined, e.g. in \cite{diehl05}.

Aim of our paper is the investigation of the model dependence of those
no-helicity flip (chiral-even) 
 GPD  of the pion upon different relativistic approaches in the spacelike 
 region, i.e. for
 negative values of $t=(p^\prime-p)^2$, where $p$ and $p^\prime$ are the
  initial and
 final four-momenta of the pion, respectively. In particular, 
  the 
  study of the GPD's in the valence and non-valence regions
 (see the following Section) is emphasized by the choice of three different
 models that explore different kinematical regions: i)
  a covariant analytic constituent quark  (CQ) model, 
that covers
the whole kinematical domain and allows us to interpolate between the other two
models; ii) two phenomenological models, elaborated within a Light-front (LF)
framework (see  
e.g.,  \cite{KP,brodsky,karmarev} for a review), which have
 a  smaller kinematical range of applicability, namely one addresses the non-valence
 region and the other the valence one.

The first model is   analytic and  covariant, and  depends upon the
mass of the constituents and a parameter, fixed by the decay constant of 
the pion. The main ingredients of such an approach are: i) the Bethe-Salpeter 
amplitude (BSA) of the pion, modeled through an analytic Ansatz in the Minkowski
space, ii) the Mandelstam formula \cite{mandel} (or Impulse Approximation
formula) for the matrix elements of the current
operator, and iii) a bare quark-photon vertex.
A peculiar feature of our Ansatz for the pion BSA  is given by the 
symmetry under the exchange of the constituent
momenta. A first version of such a model was adopted in Ref. \cite{pach02} to
investigate the frame dependence of the description of the electromagnetic
(em) pion form factor, putting in evidence the possibility to study 
the non-valence
content of the pion by using a suitable reference frame. In the present work, 
we  consider a natural extension of the model, that features a better
end-point behavior of the BSA, as well.

A second model, developed within  the LF Dynamics and already applied to the em 
pion form factor in both the space-
and timelike regions  \cite{DFPS}, is still based  on the Mandelstam formula. 
However, this model retains only the
analytic structure given by the  poles of the Dirac propagators in
 the analytic
 integration over $k^-=k^0-k^3$, i.e. the minus component of the constituent 
 four-momentum appearing in the loop formula.
An important consequence of the $k^-$-integration can be reached in a
 frame where the plus component of the virtual-photon four-momentum is different
 from zero, i.e. $\Delta^+=\Delta^0+\Delta^3\ne 0$. Indeed, in this frame 
  the contributions in the valence and non-valence regions can be
 obtained,
 allowing an investigation of the Fock components of the hadronic state
 (see \cite{brodsky,karmarev,diehlpr, hwang,diehl00} for 
 an overview of  the Fock expansion of a hadron state, within 
the LF framework). Another relevant feature of this model, that 
has a
fundamental impact in the timelike region, is the quark-photon vertex dressed
by a microscopic version of the vector meson model (VMD) \cite{DFPS}.
Finally, as explained in detail in \cite{DFPS}, the model lives in the
non-valence region, in the limit of vasnishing pion.

A third model is constructed within the 
LF Relativistic Hamiltonian Dynamics (LFHD), where the Poincar\'e covariance is 
fully satisfied
(see, e.g. \cite{KP} for a detailed review). In particular the rotational
covariance is fulfilled through the introduction of the Melosh rotations and 
the
proper definition of the   total intrinsic angular momentum. At the present
stage, the model explores only the valence region.

The paper is organized as follows: in Sec. \ref{formalism} a brief resum\'e of 
the
general properties of the pion isospin-dependent GPD's is presented; 
in Sec. \ref{Fock}, the Fock decomposition of the GPD's  is  discussed, in
view of a frame-dependent analysis;  in Sec. \ref{SYM} a covariant CQ model,
that allows an analytic evaluation of the pion GPD's is described;
in Sec. \ref{LFMan}, a first CQ Light-front model, with a quark-photon vertex
dressed by a microscopic version of the vector meson dominance model, is
presented; in Sec. \ref{LFHD}, the LFHD model, where the full Poincar\'e
covariance is implemented  is described. Finally in  Sec. \ref{Ris} and 
Sec. \ref{END} the results are discussed and the conclusions drawn.
\section{pion GPD's: kinematics and general formalism}
\label{formalism}
\begin{figure}[thb!]
\includegraphics[width=7.4cm]{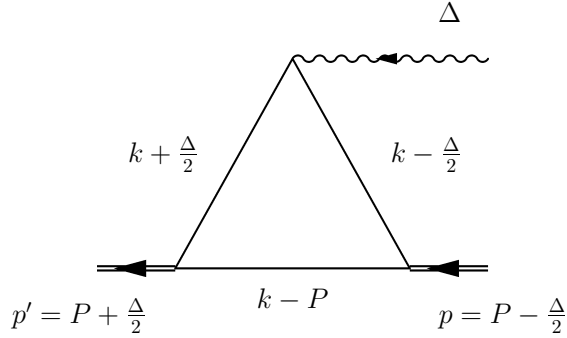}
\caption{ Diagrammatic representation of the pion GPD, with four-momenta
definitions.}
\label{fig1} 
\end{figure}

\begin{figure}[thb!]
\includegraphics[width=7.4cm]{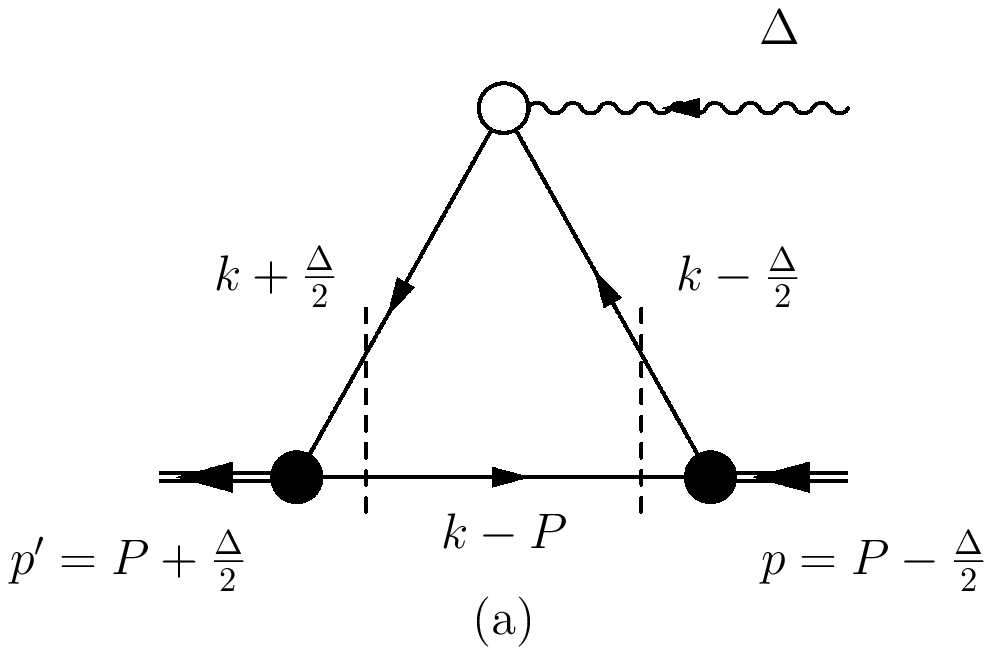}~~~
\includegraphics[width=7.4cm]{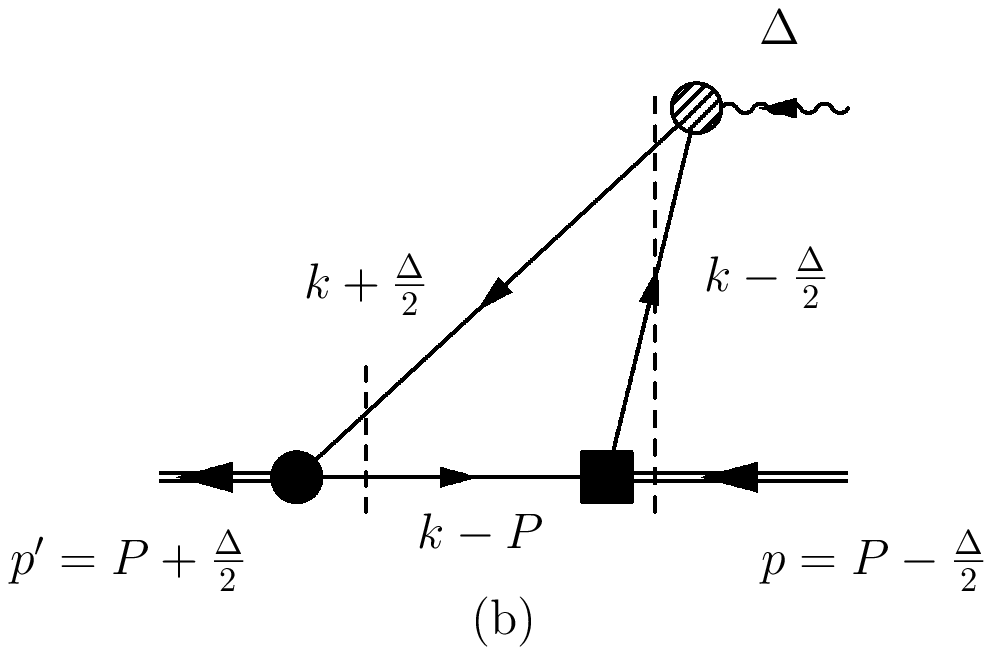}
\caption{ LF time-ordered analysis of the pion GPD.
Diagram (a): a contribution in the valence region,  $1\ge x \ge|\xi|$, 
(see text).  
Diagram (b):  a contribution in the non-valence region, 
$ |\xi|> x >-|\xi|$, (see, text).
Vertical dashed line indicate a given value of the LF time, in order to single
out the number of constituents in flight. }
\label{figbs} 
\end{figure}

In the spacelike region, let us first illustrate the kinematics of the 
DVCS
process with the symmetric momenta convention  shown in
Fig.~\ref{fig1} (see \cite{Ji}  
for the reduction of
the DVCS diagram to the one presented in Fig. 
\ref{fig1}, 
 and the pioneering paper \cite{tob92} for the  DIS regime). For on-mass-shell pions, 
i.e. $p^{\prime 2}=p^2=m_\pi^2$, and adopting 
standard notations (see, e.g. \cite{diehlpr,pasquinirev07})
\be
 t=\Delta^2=(p^\prime-p)^2,\nonu
 \xi=- \frac{\Delta\cdot n}{2~P\cdot
n}=-\frac{\Delta^+}{2~P^+}={p^+-p^{\prime +} \over p^+ + p^{\prime +}}, 
\quad \quad \quad(|\xi|\le 1),\nonu
 x=\frac{k\cdot n}{P \cdot n}=\frac{k^+}{P^+}, \quad \quad \quad (1\ge x\ge
 -1)~~,
\label{kin}\ee
where $n$ is a light-like 4-vector, such that $v^+=n\cdot v=v^0+v^3$ (the scalar
product is defined as $a\cdot b=(a^+b^-+a^-b^+)/2 -{\bf a}_\perp \cdot 
{\bf b}_\perp$),   
$P=\frac12(p'+p)$, 
and $k$ is the average
momentum of the active quark, i.e. the one that interacts with the photon (see Fig.
\ref{fig1}). Notice that  $p^+$ and $p^{\prime +}$ are necessarily positive, 
while
$\Delta^+\geq 0$ is taken 
 by choice.
From Eq. (\ref{kin})  one can trivially obtain
the following useful relations\be
p^{\prime +} = {\Delta^+ \over
2} (1- {1 \over \xi}) \quad \quad \quad
  p^{+} = ~-~{\Delta^+ \over 2} (1+{1 \over \xi})~~.
\ee
 As it is well known, the variable $x$ allows one to  single out i) the
valence region (where one has only  contributions diagonal in the Fock
 space , cf
the following Sec. \ref{Fock})
given by the union of two intervals: $x\in[-1, -|\xi|]$ (corresponding to an active
antiquark) and $x\in[|\xi|,
1]$ (corresponding to an active
quark), and ii) the non-valence region, $x\in[-|\xi|, |\xi|]$. 
In Fig. \ref{figbs}(a), it is shown  
a representative of the  contribution  with an active quark in the kinematical region $x\in[|\xi|,
1]$, (all the constituents have a plus-component of their-own momentum 
bounded from above 
by the corresponding  quantity of the parent pion).
In  Fig. \ref{figbs}(b), it is shown  a contribution 
from a pair-production process, non diagonal in the Fock space.
In Appendix \ref{kinea}  a more 
detailed kinematical discussion is given. Finally, as a short detour, let us remind that the pion 
BSA,
integrated over the minus component of the quark momentum, yields
the  two-body Fock contribution 
to the pion state,  notably non vanishing
 only in the valence sector (see \cite{karmarev}).

Within  the QCD-evolution framework, the valence region  
 is called DGLAP~\cite{dglap} region, while the non-valence one is called
 the  ERBL~\cite{erbl,lepage} region. 

In the interval $[|\xi|,1]$, the relation between the LF momentum 
fraction, $x_{q}$, of the active constituent in the initial pion 
(with the support $[0,1]$)
  and the variable $x$ defined in  Eq.
(\ref{kin}),  is given by 
\be
x_{q}={k^+ - \Delta^+/2\over p^+}={k^+ - \Delta^+/2\over P^+ - \Delta^+/2}
= {x+\xi\over 1 +\xi}={x-|\xi|\over 1 -|\xi|}~~.
\label{xb}\ee

The  isospin-dependent  GPD's 
(see, e.g. \cite{diehlpr,poly,diehl05,bronio08}) are 
 the matrix elements of
light-cone bilocal operators separated  by a light-like distance,
$z^2=z^+z^- - |{{ \bm z}_\perp}|^2=0$ ,
evaluated between pion  states with different initial and final
momenta. In
the light-cone gauge, where $A_{gluon}\cdot n=0$  and 
 the gauge link becomes unity,
  one can introduce isoscalar and
isovector combinations for the off-forward ($t \neq 0$), non-diagonal 
($\xi \neq 0$) GPD's, as follows 
\be\label{gpdisos}
  H^{I=0}_{\pi^\pm}(x,\xi,t) =  \int \frac{dz^-}{4\pi} e^{i x P^+ z^-}
\left . \langle \pi^\pm (p') | ~\bar \psi_q
(-\frac12 z) \gamma \cdot n \,  \psi_q (\frac12 z) ~| \pi^\pm (p)
\rangle \right |_{\tilde z=0}\nonu 
={1 \over 2}\left [H^u_{\pi^+}(x,\xi,t)+H^{ d}_{\pi^+}(x,\xi,t)\right]=
{1 \over 2}\left [H^{ d}_{\pi^-}(x,\xi,t)+H^u_{\pi^-}(x,\xi,t)\right]
\ee
and 
\be\label{gpdisov}
  H^{I=1}_{\pi^\pm}(x,\xi,t) = \int \frac{dz^-}{4\pi} e^{i x P^+ z^-}
~
 \left . \langle \pi^\pm (p') | \bar \psi_q~
(-\frac12 z) \gamma \cdot n ~ \tau_3 \psi_q (\frac12 z) ~
  | \pi^\pm (p) \rangle
\right |_{\tilde z=0}\nonu = {1 \over 2}\left [H^u_{\pi^\pm}(x,\xi,t)-
H^{ d}_{\pi^\pm}(x,\xi,t)\right]=\pm{1 \over 2}
\left [H^u_{\pi^+}(x,\xi,t)-H^{ d}_{\pi^+}(x,\xi,t)\right],\ee
where $\tilde z\equiv \{z^+, {\bm z}_\perp\}$,  while $\psi_q(z)$ and $\tau_3\psi_q(z)$ are the following 
doublets of  quark field
\be\left(\begin{array}{c} U(z)\\ D(z)\end{array} \right)~~, \quad \quad
\left(\begin{array}{c} U(z)\\ - D(z)\end{array} \right)~~,\ee 
respectively. 
 In Eqs. (\ref{gpdisos}) and (\ref{gpdisov}),
following \cite{poly}, instead of the Cartesian components,
 $\pi^0,\pi^1,\pi^2$, (adopted in
\cite{diehlpr,diehl05,bronio08}), 
the charged pions have been introduced, viz 
\be|\pi^\pm\rangle={|\pi^1\rangle \pm i |\pi^2\rangle \over
\sqrt{2}}~~,\qquad  \qquad |\pi^0\rangle= |\pi^3\rangle~~.\label{charged}\ee
The functions $H^u(x,\xi,t)$ and $H^ d(x,\xi,t)$ are $u$ and $d$ GPD's, 
respectively, and contain  quark and antiquark
contributions  (cf. the parton interpretation of $H^q$, e.g.,
in \cite{Ji98,diehlpr} and Fig. 2).
It is worth noting that $ H^q(x,\xi,t)$ has the support $x \in
[-1,1]$.
Finally, due to the isospin symmetry one has
\be H^u_{\pi^+}=H^d_{\pi^-},
\label{iso1}\ee
and  combining charge and
isospin symmetry (G-parity) one gets
\be {H}^u_{\pi^+}(x,\xi,t)=
-{H}^u_{\pi^-}(-x,\xi,t)= -{H}^d_{\pi^+}(-x,\xi,t)\label{iso2}
~~~.
\ee
In what follows, we deal with a charged pion and the subscript $\pi^+$ 
in the quark GPD's is dropped out whenever no ambiguity is present.
  
For vanishing $\xi$ and $t$,  one has the following partonic decomposition 
 (cf.  \cite{diehlpr,Ji98})  
 \be H^u(x,0,0)= 
\theta(x)~u(x)  - \theta(-x)~\bar u(-x)~,\nonu H^d(x,0,0)= 
\theta(x)~d(x)  - \theta(-x)~\bar d(-x)~.
\label{struc}\ee
 Equations (\ref{iso1}) and (\ref{iso2})  together with the partonic
interpretation lead to the well known relations between the standard parton
distribution functions (let us remind that the relations pertain to active
quarks), viz
\be u_{\pi^+}(x)=d_{\pi^-}(x), \quad \quad    u_{\pi^+}(x)=\bar 
d_{\pi^+}(x)~~~.
\label{iso3}\ee

The symmetry property of $H^{I=0,1}(x,\xi,t)$ (see, e.g. \cite{poly,diehlpr}) 
under the 
transformation  $x\to -x$,  that just reflects i) the charge-conjugation ($p\to -p$ and
$p^\prime \to -p^\prime$)
 and ii) the isospin invariance,    reads (reminding Eqs. (\ref{iso1}) and 
 (\ref{iso2}))
\be
H^{I=0}(x,\xi,t) ={1 \over 2}\left [{H}^u(x,\xi,t)-{H}^u(-x,\xi,t)\right]= -H^{I=0}(-x,\xi,t),\label{gpd0} \\ && 
H^{I=1}(x,\xi,t) ={1 \over 2}\left [{H}^u(x,\xi,t)+{H}^u(-x,\xi,t)\right]= H^{I=1}(-x,\xi,t))\label{gpd1}
~~.
\ee
Therefore the two GPD's are odd or even in $x$ depending upon the isospin
combination.
In addition,  under the
transformation  $\xi\to -\xi$,  that amounts
to apply  a   time-reversal transformation (since we have
to
 exchange the initial and final pion
momenta) and to exploit Hermiticity, one has (see, e.g. 
\cite{poly,diehlpr})\be
H^{I}(x,\xi,t) = H^{I}(x,-\xi,t),
\label{timrev}
\ee
namely $H^{I}(x,\xi,t)$ must be even in $\xi$.

From Eqs. (\ref{gpdisos}) and (\ref{gpdisov}) one has
\be
H^u(x,\xi,t)= H^{I=0}(x,\xi,t)+H^{I=1}(x,\xi,t), 
\nonu
H^d(x,\xi,t)=  H^{I=0}(x,\xi,t)-H^{I=1}(x,\xi,t) 
~~~.
\ee

As well known, the following sum rules hold
(note a different overall factor with respect to 
\cite{diehl05,bronio08} due to our choice of dealing with a 
charged pion,
cf Eq. (\ref{charged}))
\be
 \int_{-1}^1  dx\, {H}^{I=1}(x,\xi,t) = \int_{-1}^1  dx\, {H}^{u}(x,\xi,t)
  = ~F_\pi(t), \label{ffpi} \\ &&
 \int_{-1}^1  dx ~x ~{H}^{I=0}(x,\xi,t) = \int_{-1}^1  dx ~x ~ {H}^{u}(x,\xi,t)
  ={1 \over 2}~\left [
\theta_2(t)-\xi^2 \theta_1(t)\right ] ~~~.\label{gravff}
\ee
In Eq. (\ref{ffpi}), $F_\pi(t)$ is the pion em form factor (see Appendix
\ref{ffpiap}),
 while, according to
the Ref. \cite{Donoghue:1991qv}, in Eq. (\ref{gravff})
$\theta_1(t)$ and $\theta_2(t)$ are the  gravitational form factors (see also,
 e.g.,
\cite{poly, diehl05,bronio08b}), that enter in the parametrization of the matrix elements of
the quark part of the energy-momentum tensor (notice that in
the chiral limit
 one has $\theta_1(0) -\theta_2(0)={\cal O}(m^2_\pi) $). It should be pointed out that the
  sum rule (\ref{gravff})
for $t=0$  and $\xi=0$ yields the   longitudinal-momentum sum rule for the pion, 
i.e. $<x_q>$, as numerically illustrated in Sec. \ref{Ris}. 

For vanishing $\xi$ and $t$,  one can exploit i) Eqs. (\ref{gpd0}) 
 and (\ref{gpd1}) and ii) the partonic decomposition 
 (cf. Eq. (\ref{iso3}))  
  obtaining 
\be
{H}^{I=0}(x,0,0) ={1 \over 2}\left [{H}^{u}(x,0,0) +{H}^{d}(x,0,0)\right]={1 \over 2}\left [{H}^{u}(x,0,0)
-{H}^{u}(-x,0,0)\right]\nonu =
\theta(x)~{1 \over 2}~\left [u(x) + \bar u(x) \right ] - \theta(-x)~{1 \over 2}
\left [\bar u(-x) +u(-x)\right ]\label{struc1}~~,\ee
and 
\be
{H}^{I=1}(x,0,0) ={1 \over 2}\left [{H}^{u}(x,0,0) -{H}^{d}(x,0,0)\right]
={1 \over 2}\left [{H}^{u}(x,0,0)
+{H}^{u}(-x,0,0)\right]\nonu =
\theta(x)~{1 \over 2}~\left [u(x) - \bar u(x) \right ] - \theta(-x)~{1 \over 2}
~\left [\bar u(-x) -u(-x)\right ]
~~.
\label{pdf}\ee
Analogous relations, with   singlet, $q(x)+ \bar q(x)$, and  valence,
$q(x)- \bar q(x)$, combinations,  for the $d$-quark can be
easily obtained, by using Eq. (\ref{iso3})
 (see also \cite{poly}).  
It is worth noting that 
 for 
$~\xi=\Delta^+=0$ the ERBL region shrinks to zero and 
the variable $x$ reduces to $x_{q}$ (Eq. (\ref{xb})).  Finally,  from Eq.
(\ref{ffpi}) one  has a 
normalization for the valence combination $u_{v}(x)=u(x)-\bar u(x)$ 
given by $ \int ^1_0 dx~u_{v}(x)=1$. 

 It should be pointed out that the parton distributions  represent a bridge 
 toward the chiral-even 
transverse-momentum dependent (TMD) distribution, $f_1(x,|{\bf k}_\perp|^2)$ 
(see, e.g. \cite{jacob97,hwang08,pasqui08} for the nucleon case), as shown by the following relation  
\be
  q(x)= \int d{\bf k}_\perp~ f^q_1(x,|{\bf k}_\perp|^2), \quad \quad (x\ge 0)
\label{f1xk}
~~.\ee
Furthermore, it is worth noting that an experimental access to 
$f_1(x,|{\bf k}_\perp|^2)$
and to  other TMD's is
 a fundamental step in order to understand the correlations between
constituents inside the pion, and eventually the dynamics.

To complete this  brief resum\'e of the general formalism, 
we have to  mention that the sum rules in Eqs.
(\ref{ffpi}) and (\ref{gravff}), are the lowest order of the moments of the
isovector and isoscalar GPD's. In particular,  $H^{I=1}(x,\xi,t)$ 
(see Eq. (\ref{gpd1}))
 has only even moments, while $H^{I=0}(x,\xi,t)$  
(see Eq.
(\ref{gpd0})) has  only odd moments. Moreover, 
it turns out (see, e.g., \cite{diehlpr}) that the $n$-th Mellin moments 
of the GPD's are  polynomials of $\xi$ with
highest power $n$ for even moments and $n+1$ for odd moments, i.e. only even
powers of $\xi$ appear, as expected from Eq. (\ref{timrev}). It is worth noting
that the so-called   
{\em polynomiality}   follows from
general properties, like  Hermiticity, covariance, parity and
time-reversal invariance~\cite{Ji98,Rady01}.  The isospin-dependent moments are
given by ($j\ge0$)
\be
\int_{-1}^1 ~ dx~x^{2j} ~ {H}^{I=1}(x,\xi,t) = \sum_{i=0}^j 
A^{I=1}_{2j+1,2i}(t) (2\xi)^{2i}, \label{evenm}\\ &&
\int_{-1}^1 ~ dx~x^{2j+1} ~ {H}^{I=0}(x,\xi,t) =
\sum_{i=0}^{j+1} A^{I=0}_{2j+2,2i}(t) (2\xi)^{2i}
~~.\label{oddm}
\ee
In particular, numerical calculations of    i) $F_\pi(t)= A^{I=1}_{1,0}(t)$ and ii)
$A^{I=0}_{2,0}=\theta_2(t)/2$ and $A^{I=0}_{2,2}=-\theta_1(t)/8$, will be
presented in  Sec. \ref{Ris}. 

In conclusion,  approaches that
satisfy the basic field-theoretic assumptions  underlying
polynomiality, i.e. extended Poincar\`e covariance, automatically fulfill 
the
conditions (\ref{evenm}) and (\ref{oddm}). 
In general,  such a property is an important test of
consistency of the model.
\section{Fock decomposition}
\label{Fock}
Let us introduce the Fock expansion of the pion state, taking care of the
colorless feature of each components and including the amplitudes inside the 
kets to
simplify the notations in this Section,
(see, e.g. \cite{brodsky,karmarev}), viz
\be
|\pi \rangle =|q\bar q\rangle+|q\bar q; g\rangle +|q\bar q;q\bar q \rangle 
+ ... \quad \quad .
\ee
Then one can decompose the 
GPD's in terms of their Fock contents (see also \cite{diehlpr,hwang}), 
i.e. one can rewrite 
Eqs. (\ref{gpdisos}) and (\ref{gpdisov}) by using e.g. 
\be
H^q(x,\xi,t)=\sum_n^{{\rm Fock}}
\langle \pi; n|\Gamma^q_D|\pi;n\rangle\theta(|x|-|\xi|)\theta(1-|x|)+\nonu
+\theta(|\xi|-|x|)\left [\sum_n^{{\rm Fock}}\langle \pi; n|\Gamma^q_{ND}|\pi;n+2\rangle
\theta(\xi)
+\sum_n^{{\rm Fock}}\langle \pi; n+2|\Gamma^q_{ND}|\pi;n\rangle
\theta(- \xi)\right ]
+ \dots~\ ,
\label{fockd}\ee
where $n$ indicates the  number of quarks and antiquarks, $\Gamma_{\text D} $ and $\Gamma_{\text ND} $ are the diagonal and
non diagonal, in the Fock space, terms of the current operator, and  dots 
represent all the other
transition matrix elements, possibly containing states with gluons. The diagonal
terms  yield contributions to the valence region (DGLAP  region), while the non
diagonal ones have to  be considered in the non-valence region (ERBL region).
In Eq. (\ref{fockd}), we have shown only transitions involving fermionic 
fields, and
this explains the  selection rule $\Delta n=0,2$. 
 
 In a
simple picture of a hadron, the valence state has a dominant role
at the hadron scale, and this leads naturally to associate the DGLAP
region with this Fock component.
 
The same decomposition can be applied to the em and gravitational
form factors, and to all the $t$-dependent "generalized" form factors appearing in 
Eqs. (\ref{evenm})
and (\ref{oddm}). Clearly, this kind of decomposition could allow a deeper understanding
of the dynamics related to the components beyond the valence one.
As a simple application, let us consider the em form factor. From Eqs.
(\ref{ffpi})
and  (\ref{fockd}),  retaining only the  fermionic  transitions, 
one has 
\be\label{srval}
F_\pi^{(v)}(\xi,t)= 2\int^1_{|\xi|} dx~ H^{I=1}(x,\xi,t) 
= 2 \sum^{\text Fock}_n \int^1_{|\xi|} dx~~\langle \pi ; \;
n|\Gamma^{I=1}_{\text D} |\pi ; \; n \rangle,
\\ && \label{srnval}
F_\pi^{(nv)}(\xi,t)= 2 \int^{|\xi|}_{0} dx ~H^{I=1}(x,\xi,t)  
\nonu = 2 \sum^{\text Fock}_n \int^{|\xi|}_{0} dx~~\left[ \theta(\xi)~\langle \pi ; \;
n|\Gamma^{I=1}_{\text ND} |\pi ; \; n+2 \rangle + 
\theta(-\xi)~\langle \pi ; \;
n +2 |\Gamma^{I=1}_{\text ND} |\pi ; \; n \rangle\right ] \ .
\ee
The valence term,
$F_\pi^{(v)}(\xi,t)$, receives the largest contribution from the valence
component of the pion state,  but it  does not give the full result in
the whole kinematical range, as indicated by the residual dependence upon $\xi$.  
The non-valence term, $F_\pi^{(nv)}(\xi,t)$,  is due to  contributions 
like   the pair-production mechanism, see Fig. \ref{figbs}(b).  The sum of
Eqs. (\ref{srval}) and (\ref{srnval}) leads to the full 
result, viz
\be
F_\pi(t) =F_\pi^{(v)}(\xi,t) +F_\pi^{(nv)}(\xi,t)
\label{srtot}\ee
and it  is {\it independent of $\xi$} and function of $t$ only. One can
also express the invariance of the sum under changes of $\xi$ as:
\be
 \frac{\partial^m}{\partial \xi^m}F_\pi^{(v)}(\xi,t)
 =-\frac{\partial^m}{\partial \xi^m}F_\pi^{(nv)}(\xi,t)
\ , \label{deri}
\ee
with $m\ge 1$.
It is worth noting that all the derivatives of $F_\pi(t)$ are independent upon
$\xi$, and therefore relations like the one in Eq. (\ref{deri}) can be
generalized, i.e.
\be
 \frac{\partial^m}{\partial \xi^m}\frac{\partial^\ell}{\partial t^\ell}
 F_\pi^{(v)}(\xi,t)
 =-\frac{\partial^m}{\partial \xi^m}\frac{\partial^\ell}{\partial t^\ell}
 F_\pi^{(nv)}(\xi,t)
\ , \label{deri1}
\ee
with $m\ge 1$ and $\ell \ge 0$.
As a consequence, with the help of Eq. (\ref{ffpi}), one can deduce interesting 
sum rules for the
partial derivatives of $H^{I=1}(x,\xi,t)$.

Let us remind that calculations of  the elastic form factors have been performed
in  different frames. In particular,  it has been chosen i) the Drell-Yan frame,
 where $\Delta^+=0$  
and   therefore $\xi=0$ (see 
Ref.
\cite{brodsky} for generalities on the Drell-Yan frame), or ii) a Breit frame 
(i.e. $\Delta^+=-\Delta^-$) where ${\bm \Delta}_\perp
=0$  (see \cite{lev98} for an extended discussion of the motivations for adopting
such a frame), and then $\xi$ follows a  kinematical trajectory in the 
$(\xi,t)$-plane
given by $|\xi|=1 /\sqrt{1-4m_\pi^2/t}$ (see below Eq. (\ref{xiq})). In the first case, the em form factor
 is
saturated by 
 the valence contribution, because of $\xi=0$ (cf Eqs. (\ref{srtot}) and
 (\ref{srval})), while in
the second  frame both  valence and non-valence terms contribute, since
$\xi$ does not vanish, but changes with $t$. For $-t^2>>m^2_\pi$ the value 
of $\xi$
approaches 1, and therefore the non-valence term
saturates the em form factor (cf Eqs. (\ref{srtot}) and
 (\ref{srnval})). In model calculations this general behavior
was indeed observed~\cite{pach02}. It is understood that for an experimental
investigation of the
whole $(\xi,t)$-plane, different kinematical conditions are needed, also
exploiting the helpful properties of the LF boosts (see, e.g. \cite{KP}).

Following the same spirit, one could extend this analysis  to the other 
form factors that
appear in Eqs. (\ref{evenm}) and (\ref{oddm}), i.e. one can  consider the
 partial
derivatives of the valence and non-valence contribution to the generalized form
factors $A^{I=0}_{2j+2,2i}(t)$ and $A^{I=1}_{2j+1,2i}(t)$, obtaining   final  relations  
 that have  the same structure as the
ones in Eqs. (\ref{deri}) and (\ref{deri1}).

\section{  Covariant Model of the pion with Pauli-Villars regulators }
\label{SYM}
 In Ref. \cite{pach02}, an
analytic covariant model, symmetric in the exchange of the constituent
four-momenta (see  Refs. \cite{pach99,bakker01} for previous non
symmetric versions) was adopted 
for evaluating the em form factor. In this work,  a direct
extension of the symmetric covariant model to DVCS is  exploited for calculating the
no-helicity flip GPD, in the spacelike interval  $0 \ge ~t~\ge -10$ (GeV/c)$^2$. 

 In a Breit frame, one has  $\Delta^0=0$, i.e. $\Delta^+=-\Delta^-$, and 
${\bf p}^\prime=-{\bf p}= {\bf  \Delta}/2$. By choosing  
  $\Delta^+ \ge 0$,  and reminding that 
 \be
p^{\prime -} ={m^2_\pi+|{\bm \Delta}_\perp|^2 /4\over p^{\prime +}}, 
\quad \quad
p^{ -} ={m^2_\pi+|{\bm \Delta}_\perp|^2/4 \over p^{ +}},\nonu
\Delta^-=p^{\prime -}-p^{ -}=-\Delta^+~
{m^2_\pi+|{\bm \Delta}_\perp|^2/4\over p^{\prime +}p^{ +}}=
\Delta^-~
{m^2_\pi+|{\bm \Delta}_\perp|^2/4\over (p^{ +}+\Delta^+)p^{ +}},
\ee 
  one gets 
\be
p^{+} =
  {-\Delta^+ + \sqrt{-\Delta^2 +4  m^2_\pi}\over 2}= ~-~{\Delta^+ \over
2} (1+{1 \over \xi}),\nonu 
p^{\prime +} ={\Delta^+ + \sqrt{-\Delta^2 +4  m^2_\pi}\over 2}= 
{\Delta^+ \over
2} (1- {1 \over \xi})~~~.
\ee
Then the   following relation  holds (notice that $2~P^+=\sqrt{-\Delta^2 +4  m^2_\pi}$)
\be
\Delta^2=-\Delta^{+2}-|{\bm \Delta}_\perp|^2=
-4 \xi^2 P^{+2}-|{\bm \Delta}_\perp|^2=
- \xi^2~\left(-\Delta^2 +4  m^2_\pi\right)-|{\bm \Delta}_\perp|^2 ,
\ee
which leads to a constraint on the
maximal value for the variable $\xi$. As a matter of fact,
 in the spacelike region $-\Delta^2 +4 
m^2_\pi\ne 0$ and one has
\be
\xi^2= {-\Delta^2 -|{\bm \Delta}_\perp|^2\over -\Delta^2 +4 
m^2_\pi}\ .
\label{xiq}\ee
Then, the  maximum value of $\xi^2$ is 
found for ${\bm \Delta}_\perp=0$, viz 
\be
\xi^2\le {-\Delta^2 \over -\Delta^2 +4 
m^2_\pi}\leq 1 \ .
\ee
For $m_\pi=0$ (and $\Delta^2\neq 0$), one has \be
\xi^2=1+{|{\bm \Delta}_\perp|^2\over \Delta^2 }\ .
\label{nompi}\ee If one  additionally chooses a frame where 
${\bm \Delta}_\perp=0$ 
(i.e. only $\Delta_z\ne 0$), then $\xi=-1$  and therefore,
 in this extreme case,
  only the
non-valence region contributes.

A basic ingredient in the analytic covariant model of Ref. \cite{pach02} is the pion BS amplitude,
that can be quite well approximated by retaining only the pseudo-scalar
 Dirac structure (see, e.g., \cite{maris}), namely 
\be
\Psi (k-P,p) =  -{m \over f_\pi}~S\left(k-\Delta/ 2\right)~  \gamma^5 ~
\Lambda(k-P,p)~
S\left(k-P\right), \label{bsa}
\ee
where ${m \over f_\pi}$ is the quark-pion coupling, as suggested by a simple 
effective Lagrangian (see,
e.g. \cite{tob92}),  $f_\pi=92.4$ MeV   the pion decay 
constant,   $m$  and  $S(k)$ are the mass and  
  the Dirac propagator of the constituent quark (CQ), respectively. 
  In Eq. (\ref{bsa}), $\Lambda(k-P,p)$ is a scalar
function that describes
 the momentum-dependent part of the coupling between the
constituents and the spin-0 system and plays the role of the Pauli-Villars
regulator of the otherwise divergent integrals that yield  GPD's
 or    the em form
factor.   In particular in this work we adopt two
symmetric (in the exchange of the CQ four-momenta) covariant 
forms: i) the one considered in Ref. \cite{pach02}, and based on the following sum
\be
 \Lambda_1(k-P,p)= ~C_1~\left \{
\frac{1}{\left[\left(k-\Delta/2\right)^2-m^2_{R} + \imath \epsilon\right]}+
\frac{1}{\left[\left(P-k \right)^2-m^2_{R}+ \imath \epsilon\right]}\right\},
\label{vertexs}
\ee
and ii) a natural extension based on a product, viz
\be
 \Lambda_2(k-P,p)= ~C_2~ 
\frac{1}{\left[\left(k-\Delta/2\right)^2-m^2_{R} + \imath \epsilon\right]}~
\frac{1}{\left[\left(P-k \right)^2-m^2_{R}+ \imath \epsilon\right]} \ .
\label{vertexp}
\ee
This product-form provides a more realistic transverse-momentum fall-off, as
seen from the expected behavior of the BS amplitude obtained by using a simple
(one-boson-exchange) kernel (see, e.g., \cite{carbonellepja})), and this has a
sizable
impact on both the high-momentum tail of the em form factor and the end-point
behavior of the parton distribution, as shown in the results presented in Sec.
\ref{Ris}. We can anticipate that the  most favorable comparison with the
experimental data of the em form factor is obtained by using the
product-form, as also expected if one follows a  pQCD analysis, where
  a  one-gluon exchange represents the leading contribution to
  the kernel \cite{lepage,sawi85}.
 
In both expressions, once the constituent mass $m$ is chosen, $m_R$ is determined by fitting the experimental value for
$f_\pi$ (cf \cite{pach02}), while the constants $C_1$ and $C_2$ are fixed 
by exploiting the charge normalization, as discussed below.

As a final comment on the Dirac structure that appears in Eq. (\ref{bsa}), we
remind that it leads to the standard Melosh rotation for a pair of fermions
coupled to a total spin $S=0$ (see \cite{jaus}), once we consider 
 the valence wave
function, defined as follows (see, e.g. \cite{karmarev})
\be
\Phi_{val}(\kappa^+,{\bm \kappa}_\perp,p)=-{m \over f_\pi}~
\int {d\kappa^-\over 2 \pi}~
S_{on}\left(\kappa-p\right)~  \gamma^5 ~
\Lambda(\kappa,p)~
S_{on}\left(-\kappa\right)
\label{phival}\ee
where $S_{on}(\kappa)=(\psla \kappa_{on}+m)/(\kappa^2-m^2+i\epsilon)$ with
$\kappa^\mu_{on} \equiv \{\kappa^-_{on}=(m^2+|{\bm \kappa}_\perp|^2)/ \kappa^+,
\kappa^+,{\bm \kappa}_\perp \}$.

The no-helicity flip GPD  for  the pion is calculated in one-loop
approximation (triangle diagram cf. Fig. \ref{fig1}) with the
BS amplitude of  Eq. (\ref{bsa}) and the symmetrical forms shown 
in Eqs. (\ref{vertexs}) and 
(\ref{vertexp}). In particular, the $u$-quark GPD 
 is given in Impulse Approximation by
\be
 H^u(x,\xi,t) = -\imath ~N_c~{\cal R}\nonu
\times\int
\frac{d^4k}{2(2\pi)^4}\delta(P^+x-k^+) \; V^+(k,p,p^\prime) 
 \Lambda(k-P,p^{\prime})\; \;
\Lambda(k-P,p) \ , \label{jmu}
\ee
where $N_c=3$ is the number of colors, ${\cal R}=2  m^2/f^2_\pi$  and 
\be
 V^+(k,p,p^\prime)=Tr\left \{ S\left({k}-{P}\right)
\gamma^5 ~S\left({k}+\frac{
{\Delta}}{2}\right) 
\gamma^+~S\left({k}-\frac{{\Delta}}{2}\right)\gamma^5\right \}~
\ . \label{trace}
\ee
The presence of the delta-function in Eq. (\ref{jmu}), given the kinematical 
relations in
Eq.  (\ref{kin}), imposes the correct support $[-|\xi|,1]$ for the variable $x$
as discussed in details in Appendix \ref{intkm} (note that  $H^d$ has the support
$[-1,|\xi|]$ for the variable $x$) .
 A relevant feature in the analysis of the GPD, as well as in the case of the em
  form factor, is  given by
  the
instantaneous term present in $S(k)$. As a matter of
fact,  the Dirac propagator can be decomposed using the LF kinematics
 as follows \cite{brodsky}
\be
S(k)=\frac{\psla{k}+m}{k^2-m^2+\imath \epsilon}=S_{on}(k)+\frac{\gamma^+}{2k^+}
=\frac{\psla{k}_{on}+m}{k^+(k^--k^-_{on}+\frac{\imath
\epsilon}{k^+})} +\frac{\gamma^+}{2k^+} \ , \label{inst}
\ee
where the second term, proportional to $\gamma^+$, is an
instantaneous one in the LF time. It should be pointed out
that the instantaneous contribution to the GPD is produced
only by the spectator fermion (in the present example an antifermion), i.e. by 
$S(k-P)$. Indeed, the instantaneous terms
pertaining to the other propagators do not contribute, because  of
 the property $(\gamma^{+
})^2=0$. In our symmetric model, the instantaneous term of Eq. (\ref{inst})
contributes to $H^u(x,\xi,t)$ both in the valence and in the non-valence region
 (see Eqs. (\ref{i1})-(\ref{b5p})),
 since we take fully into account  the analytic structure of the
symmetric vertex function (for a different approach, where such an 
analytic structure is disregarded see \cite{Ji}). 

The pion em form factor
is  obtained by using the sum rule (\ref{ffpi}):
\be
F_\pi(t)=~\int^1_{-1} dx ~H^u(x,\xi,t)\nonu =  -\imath N_c~{{\cal R}\over 
( p^{\prime +}+p^+)}
\int
\frac{d^4k}{(2\pi)^4} \; V^+(k,p,p^\prime) 
 \Lambda(k-P,p^{\prime})~\Lambda(k-P,p)\ .
 \label{ffpimandel}
\ee
The last expression  for   $F_\pi(t)$ can be extracted directly from  the Mandelstam formula~
for the matrix elements of the em current \cite{mandel} (see,
e.g., \cite{Bro69}, \cite{Riska94}), as well. Notice that    
the model preserves current conservation, as discussed in \cite{pach02}.

The normalization of the form factor, Eq. (\ref{ffpimandel}), allows us to
determine   $C_1$ and $C_2$  in Eqs. (\ref{vertexs}) and (\ref{vertexp}). 
Such a
charge normalization represents
 the impulse approximation of the
normalization condition in the fully interacting BS theory \cite{mandel,lurie}.

A standard analytic integration on $k^-$ (see Appendix \ref{intkm} for details) 
leads to the
following decomposition of $H^u(x,\xi,t)$ in valence and non-valence 
contributions
\be
H^u(x,\xi,t)=
 H_{(v)}^u(x,\xi,t)\theta(x-|\xi|)\theta(1-x) 
 +H_{(nv)}^u(x,\xi,t)\theta(|\xi|-x)\theta(|\xi|+x)\label{jmu1}\ .\ee
Notice that    $H_{(v)}^u$ and $H_{(nv)}^u$ are given in Appendix \ref{intkm}
  for the two momentum dependences  shown in Eqs. (\ref{vertexs}) and  
  (\ref{vertexp}). 

 The $d$-quark GPD can be obtained reminding Eq. (\ref{iso2}).
   
 Within our covariant model the  valence component $H_{(v)}^u$ in Eq.
 (\ref{jmu1}) is an approximation to the diagonal terms in  Eq.
 (\ref{fockd}), while the component $H_{(nv)}^u$ contains
the contribution of the pair-production mechanism from an incoming
virtual photon with $\Delta^+ \ > \ 0$ and approximates the non diagonal terms.

 An interesting approximation of the  contribution to GPD in the valence
region can be
obtained once the analytic structure of the BS amplitude is disregarded and 
only the poles of the propagators are retained in the integration over $k^-$ (see
Appendix \ref{intkm}). As a matter of fact, see Eq. (\ref{hon}), within the
mentioned approximation
\be
H_{(v)}^u(x,\xi,t)\sim H_{(v)on}^u(x,\xi,t)=
  - ~{N_c~{\cal R}\over 4(2\pi)^3}~
\int d \bm \kappa_{\perp} \int^{p^+}_0 d \kappa^{+} 
{\delta\left[P^+(1-x)-\kappa^+\right] \over \kappa^+(p^+-\kappa^+)
 (p^{^{\prime}+}-\kappa^+)} \nonu 
\times~Tr[{\cal O}^+(\kappa^-_{on})]
~  
\frac{\left. \Lambda(\kappa,p)\right|_{\kappa^-_{on}}}
{ \left[p^- - \kappa^-_{on} -(p-\kappa)^-_{on}\right]}~
\frac{\left. \Lambda(\kappa,p^\prime)\right|_{\kappa^-_{on}}}
{\left[p^{\prime -} - \kappa^-_{on} -(p^\prime-\kappa)^-_{on}\right]},
\label{hon2}\ee
where $\kappa=P-k$,  $\kappa^-_{on}=(m^2+|{\bm \kappa}_\perp|^2)/ \kappa^+$ and 
\be
Tr[{\cal O}^+(\kappa^-_{on})]=Tr\left \{ \left(\psla{\kappa}_{on}+m\right)
\left[ \left(\psla p^\prime-\psla{\kappa}\right )_{on}+m\right]~ 
\gamma^+~\left[\left(\psla p-\psla{\kappa}\right )_{on}+m\right]\right \}\ .
\ee
Moreover, if  in Eq. (\ref{hon2}) we identify the following ratio  
$$
\frac{\left. \Lambda(\kappa,p)\right|_{\kappa_{on}}}
{ \left[p^- - \kappa^-_{on} -(p-\kappa)^-_{on}\right]}
$$
 with a model LF wave function, then the final expression 
coincides with the result  obtained within a LFHD approach (see the
following Sec \ref{LFHD}), since the   trace $Tr[{\cal O}^+(\kappa^-_{on})]$
generates the correct Melosh-rotation factor \cite{jaus}. We would stress that
the identification is meaningful once the analytic
structure of the BS amplitude is disregarded.

\section{ Light-front Models of the pion}
\label{LFM}
In this Section we  present models that at different extent i) fulfill the
Poincar\'e covariance  and ii) take into account the Fock components of the pion
state beyond the valence contribution. A first important difference between 
the
models is given by the  frame we choose. In the approach we call  
Mandelstam-inspired
LF model, a Breit frame, where ${\bm\Delta}_\perp=0$, is considered. This choice
was followed in Ref. \cite{DFPS} in order to perform a microscopical calculation
of the em pion form factor in both the space- and timelike regions. It should be
pointed out that such a frame leads to consider contributions from a
pair-production mechanism, differently from what happens in a Drell-Yan frame,
where  $\Delta^+=0$. This second frame is the one adopted in  the
second approach illustrated in this Section, based on a LF Hamiltonian Dynamics
 description of the pion state (see, e.g. \cite{KP} for a general review
of LFHD).
\subsection{Mandelstam-inspired LF Model}
\label{LFMan}
In Ref. \cite{DFPS} an approach was elaborated to calculate the em form
factor of the pion starting from a covariant expression of  the matrix
elements of the
 current given by  
the  Mandelstam formula \cite{mandel} (cf also Eq. (\ref{jmu})). Moreover,   a
microscopic VMD was used for dressing the quark-photon vertex. The dynamical
inputs of such an approach were the wave functions of both the pion and vector
mesons, taken as  eigenstates  of the relativistic CQ square
mass operator of Ref. \cite{FPZ}, 
which includes both
confinement, through a harmonic oscillator potential, and  
$\pi-\rho$ splitting through a Dirac-delta interaction in the pseudoscalar 
channel. In what follows we apply the same approach for evaluating the no-helicity
flip GPD's.

Let us first illustrate the kinematics in the adopted frame, where 
${\bm \Delta}_\perp=0$ (i.e. 
$\Delta^-={\Delta^2 / \Delta^+}$) and  ${\bm
p}_\perp={\bm p}^\prime_\perp=0$. Then 
in the spacelike region, for $\Delta^+ \ge0$, one has for $p^{+}$ and 
$p^{\prime+}$ 
\be 
  p^{+} ={\Delta^+\over 2}~\left(-1 + \sqrt{1-4  {m^2_\pi \over\Delta^2}}\right)
  = ~-~{\Delta^+ \over
2} (1+{1 \over \xi}),\nonu
p^{\prime +} ={\Delta^+\over 2}~\left(1 + \sqrt{1-4  {m^2_\pi
\over\Delta^2}}\right)
= {\Delta^+ \over
2} (1- {1 \over \xi}),
\ee
since
\be
p^{\prime -} ={m^2_\pi \over p^{\prime +}}, \quad \quad
p^{ -} ={m^2_\pi \over p^{ +}},\nonu
\Delta^-=p^{\prime -}-p^{ -}=-\Delta^+~
{m^2_\pi\over (p^{ +}+\Delta^+)~p^{ +}}\ .
\ee
The following simple relation between $\xi$ and $\Delta^2$ holds
\be
\xi=-{\Delta^+ \over 2 P^+}=-{\Delta^+ \over(p^{\prime +}+ p^{ +})}=-{1 \over \sqrt{1-4  {m^2_\pi
\over\Delta^2}}}\ .
\label{xilfm}\ee
It is easily seen that if  $m_\pi=0$ one has $\xi=-1$ for any $\Delta^2$.

Extending the approach of Ref. \cite{DFPS}, one can find for the quark GPD 
  the same formal expression of Eq. (\ref{jmu}), but
i)  a microscopic VMD dressing, 
$\Gamma^{\mu}(k,\Delta)$,  is 
considered instead of the bare quark-photon vertex, $\gamma^\mu$, and
 ii) phenomenological 
Ansatzes
for the BS amplitudes in the valence and non-valence regions are adopted. 
Another basic
difference with respect to the analytic model presented in the previous 
Section, is that only the simple 
 analytic structure of the Dirac propagators is retained, i.e.  the
 analytic structure  is disregarded in the BS amplitudes of both i) the initial and final
  pion and ii)
 the  VM dressing of the quark-photon vertex.
   This approximation 
 turns out to be a very effective one in the calculation of the em form 
 factor just 
 in the ${\bm \Delta}_\perp=0 $  
 frame \cite{ppa}. 

 In Ref. \cite{DFPS}, a further
 simplification in the calculation was achieved by a quite natural assumption, 
 namely
a vanishing pion mass. Within such an approximation 
only diagrams with a $q \bar q$ production  contribute (cf Fig. \ref{figbs}(b)), 
and this
implies the necessity to introduce  
 the VMD dressing. We have to stress  that a bare term is missing, 
due to the vanishing pion mass (cf the discussion in \cite{DFPS}).  Therefore,
in   the quark-photon vertex for the covariant model,  Eq. (\ref{trace}),
the Dirac matrix $\gamma^+$ is replaced by
 the plus component of the following four-vector, that  microscopically describes
 a VM dressing. For $t\le~0$ one has  
 \be
 \Gamma^{\mu}(k,\Delta) =\sqrt{2} \sum_{n, \lambda}~
\left [ \epsilon_{\lambda} \cdot \widehat{V}_{n}(k,P_n)  \right ]  ~ 
\Lambda_{n}(k,P_n) ~ 
{  [\epsilon ^{\mu}_{\lambda}]^* {f_{Vn} }\over (t -
M^2_n )} \ , 
\label{VMD}\ee
where
$f_{Vn}$ is the decay constant of the n-th VM
 into a virtual photon ({\em calculated in the model}),  $P^\mu_n\equiv
 \{M^2_n/\Delta^+,\Delta^+ ,{\bf 0}_\perp\}$  the
 four-momentum of an on-mass-shell VM with a square mass given by
 $P^2_n=M^2_n$ 
 and
 $\epsilon_{\lambda} (P_n)$ its polarization. Moreover, the VM BS amplitude is
 approximated as follows     
\be
\Psi_{n \lambda} (k,P_n) =
\frac{\psla{k} + m}{k^2 - m^2 + \imath \epsilon}
\left [ \epsilon_{\lambda}(P_n) \cdot \widehat{V}_{n}(k,P_n)  \right ]  ~
\Lambda_n(k,P_n) ~
\frac{\psla{k}-\rlap\slash{P_n} + m}{(k - P_n)^2 - m^2 + \imath
\epsilon} \ , 
\label{bsan}
\ee
where $\widehat{V}_{n}(k,P_n)$ is the proper Dirac structure, and 
$\Lambda_{n}(k,P_n)$ the momentum-dependent part, 
approximated on the LF hyperplane, as discussed
below.
 
In the valence sector,  after performing the $k^-$ integration,  
 both pion and VM's  BS amplitudes reduce to
 3D  amplitudes with one constituent 
on its mass shell. In  \cite{DFPS}, the  momentum-dependent part 
of the on-shell 
 VM BS amplitude  (that contains on both sides proper Dirac projectors) is described through 
 a LF  VM wave function, i.e.
\be 
\frac{ P^+_{n} ~ \left .\Lambda_{n}(k,P_n) \right | _{k^-=k^-_{on}}}
{[M^2_n - M^2_0(k^+, {\bf k}_{\perp}; P^+_{n})]} =
\psi_{n}(k^+, {\bf k}_{\perp}; P^+_{n}),  ~   
\label{bspion}\ee 
and  $$M^2_0(k^+, {\bf k}_{\perp}; P^+_{n})]= P^+_n~
\left [ k^-_{on} +\left (P_n -k \right )^-_{on}
\right ]~~~.$$ 
 In Eq. (\ref{bspion}), 
$\psi_{n}(k^+, {\bf k}_{\perp}; P^+_n) ~ $ is 
an
 eigenfunction of the relativistic CQ square
mass operator of Ref. \cite{FPZ}, as mentioned at the beginning of this Section.
 Moreover, it is  normalized to the probability
of the valence Fock state, according to the model elaborated in 
\cite{DFPS}. 

The valence component of the pion was modeled adopting an analogous Ansatz.
 Moreover, in \cite{DFPS} two different calculations were generated by 
using i) the pion eigenstate
 of the model in Ref. \cite{FPZ} and ii) the pQCD asymptotic wave function (see,
 e.g. \cite{lepage}).

In the non-valence region, namely the only region contributing to
the GPD's for $m_\pi=0$ (see Eq. (\ref{xilfm})),  besides the pion valence 
component in the initial state one has to
deal with  
a non-valence component of the pion state, since the process depicted in Fig.
\ref{figbs}(b) can be interpreted  as  a transition from  a state composed by the 
valence component of the initial pion and the virtual photon, $ |q\bar q,
\gamma^*\rangle$, to  a higher Fock component, $  |q\bar q, q \bar q
\rangle$, pertaining to the final pion.  At level of the pion BS amplitudes, one has to model 
an off-shell  BS amplitude, that takes into account the absorption of the 
initial
 pion
 by an antiquark (according to the case illustrated in Fig. \ref{figbs}(b)).  
 In \cite{DFPS} a simple Ansatz, namely a constant vertex  was assumed, 
 like in 
 Ref. \cite{Choi}.  Notice that  such a 
coupling constant is fixed by the normalization of the pion form factor, since the
diagram shown in Fig. \ref{figbs}(a) does not contribute, as 
 a consequence of the simplification $m_\pi=0$.

Within the approach presented in this subsection, since $|\xi|=1$, 
(given the vanishing $m_\pi$) the quark GPD has only
contribution from $H^u_{(nv)}$, i.e. 
\be
H^u(x,|\xi|=1,t)=  H_{(nv)}^u(x,|\xi|=1,t)~~\theta(1-x)~\theta(1+x),
\ee
where, introducing $\kappa =P-k$, 
\be
H_{(nv)}^u(x,|\xi|=1,t)= ~-~\sum_n~{f_{Vn} \over t -M^2_n}~ 
 {N_c \over (2 \pi)^3} 
{{\cal{D}}_\pi \over \sqrt{2} } \int_{p^+}^{p^{\prime +}} {d\kappa^+ ~\delta
\left [P^+(1-x) -\kappa^+\right ]
\over \kappa^+ ~ (p^{\prime +} - \kappa^+) ~ (p^{ +} - \kappa^+)
} \int d{\bm \kappa}_{\perp} ~\times   
\nonu   
\left \{ { \psi_{n}((p^\prime -  \kappa )^+, -{\bm \kappa}_{\perp}; P^+_{n}) ~ ~
\left[M^2_n - M^2_0(\kappa^+, {\bm \kappa}_{\perp}; P^+_{n})\right] \over 
\left [ t-M^2_0(\kappa^+, {\bm \kappa}_{\perp}; P^+_{n}) + i\epsilon \right ]
} ~{\cal I}_{1} ~ + 
\psi^* _{\pi}((p^\prime -  \kappa )^+, -{\bm \kappa}_{\perp}; p^{\prime +}) ~ ~ 
{\cal I}_{2}  \right \}
\quad  \ , 
\ee
where ${\cal{D}}_\pi$ is the constant describing the off-shell quark-pion
vertex, while ${\cal I}_1$ and ${\cal I}_2$ are given by 
\be
 {\cal I}_{1} = ~-
 \frac{1} { 2 } ~ \frac{m}{f_\pi} ~
 \left . \Lambda((p^\prime -  \kappa ), p^\prime)
 \right | _{\kappa^-=p^{\prime -} - (p^\prime -  \kappa)^-_{on}}
  \nonu 
 \times~Tr \left \{ \gamma^+ ~  [(\psla p^\prime - \psla \kappa)_{on} + m] ~ 
\left [ \widehat{V}_{nz}(p^\prime -\kappa, P_n) ~ \right ] _{on} 
~[(\psla p - \psla \kappa)_{on} + m]  \right \} ,
\label{T1IInbb}  
\nonu
{\cal I}_{2} = \frac{1} { 2 }   
 Tr \left \{(\psla \kappa_{on} + m)
 [(\psla p^\prime - \psla \kappa)_{on} + m] 
\left [ \widehat{V}_{nz}(p^\prime -\kappa, P_n) ~ \right ] _{on} 
 \gamma^+  \right \}\nonu \times \left . \Lambda_{n}(p^\prime -\kappa,P_n) \right | _{\kappa^-=p^{\prime -} 
- (p^\prime -  \kappa)^-_{on}} \ .\nonu
\label{T3IInbb}
\ee
The Dirac structure, 
$\left [\widehat{V}^{\mu}_{n}(p^\prime -\kappa,P_n)\right ]_{on}$, where all the constituents
are  on their own mass-shell, is chosen 
in order to 
generate the proper Melosh rotations for $^3S_1$ states \cite{jaus}. 
Furthermore,  the  traces previously shown contain  the
 instantaneous terms  (see Eq. (\ref{inst})) that survive after assuming
 $m_\pi=0$.
 In order to model  the instantaneous part of the vertex functions directly
 attached to   $\gamma^+$, we performed the following replacements
 \be
 {m \over f_\pi}\left .\Lambda((p^\prime -  \kappa ), p^\prime)
 \right | _{\kappa^-=p^{\prime -} - (p^\prime -  \kappa)^-_{on}}
\to {\cal C}_\pi~
\psi_{\pi}(\kappa^+, {\bm \kappa}_{\perp}; p^{\prime +})~
{[m^2_\pi - M^2_0(\kappa^+, {\bm \kappa}_{\perp}; p^{\prime +})] 
\over p^{\prime +}}\label{piinst}\ee
for the pion, and 
 \be\left .\Lambda_{n}((p^\prime -  \kappa ),P_n) \right | _{\kappa^-=p^{\prime -} - 
 (p^\prime -  \kappa)^-_{on}}\to\nonu  {\cal C}_{VM}~
 \psi_{n}((p^\prime -  \kappa )^+, -{\bm \kappa}_{\perp}; P^+_{n})~
{[M^2_n - M^2_0((p^\prime -  \kappa )^+, -{\bm \kappa}_{\perp}; P^+_{n})] \over P^+_n}
 \label{vminst}\ee for the VM's, as in
  \cite{DFPS}. 
 In Eqs. (\ref{piinst}) and (\ref{vminst}), 
the constants ${\cal C}_\pi$ and  ${\cal C}_{VM}$ roughly describes the 
effects of the 
short-range interaction. Indeed, 
  a relative weight, $w_{VM} = {\cal C}_{VM} / {\cal C}_{\pi}$, can be used as 
a free parameter. Let us remind that the on-shell part of the BS amplitudes have
on the left and right sides  the proper Dirac projectors.

Finally, it is worth noting that the results presented in the following 
Section \ref{Ris} have been
calculated by using  all the
parameters adopted in  \cite{DFPS}, but
with
a CQ mass  $m=200$ MeV and $w_{VM}=-1$ (see   \cite{DFPS}  
for $m=265$ MeV and  different values for $w_{VM}$). It should be pointed out
 that only one adjusted parameter is necessary for describing the em form factor
 in
the spacelike region.

The model remains invariant for kinematical transformation, after the
approximation we have applied.

\subsection{Light-front Hamiltonian
Dynamics model}
\label{LFHD}
Within a LFHD approach (see \cite{KP} for a review of 
 the three
forms of the relativistic HD introduced by Dirac in \cite{dirac}) 
the Poincar\'e
covariance  of the description of the pion can be fully implemented, once 
the current operator is chosen in
order to fulfill the proper commutation rules with respect  to all the
generators (i.e. both  the kinematical and the dynamical ones). 
A widely adopted strategy, within the LFHD
approach, is to model the em current by using a one-body operator, but 
in the
Drell-Yan frame, i.e. where $\Delta^+=0$. For instance, in this 
frame the em form factor 
can be
obtained by using only the matrix elements of the plus component of the current
operator, and this allows to overcome some difficulties that manifestly  
appear  for
hadrons with angular momentum $\ge~1$ (see \cite{KP} and \cite{lev98} for 
a general discussion).

In the Drell-Yan frame, ${\bm \Delta}_\perp \ne 0$ and one can choose ${\bm p}^\prime_\perp=
-{\bm p}_\perp={\bm \Delta}_\perp/2$. It is worth noting that  
only  the spacelike region can be addressed, since  $\Delta^2 = - 
|{\bm \Delta}_\perp|^2$.  Moreover, one has  $p^{+}=p^{\prime+}$  and therefore
$\xi=0$ for any $\Delta^2$.

In this section, the LFHD model with CQ's, already successfully applied for
describing the charge form factor and decay constant of the 
pion~\cite{Chung:1988mu,Schlumpf:1994bc}, is adopted for investigating
  the DGLAP contribution to the no-helicity flip GPD.
This corresponds to consider in the Fock-space expansion of Eq. (\ref{fockd})
the diagonal contribution with $n=2$ constituents (i.e. the valence
component, cf also Eq. (\ref{phival}), introducing 
the explicit representation in terms of overlap of light-cone 
wave functions (LCWFs) \cite{brodsky,hwang}.
The quark contribution
to the  GPD in the region  $0\leq x\leq 1$ 
can be written in terms of the LCWF $\Psi_\pi(x,{\bm \kappa}_{\perp};\lambda_q,\lambda_{\bar q})$ 
for the quark-antiquark system
as
\be
H^u(x,\xi=0,t)  = 
\sum_{\lambda^\prime_q,\lambda_q,\lambda_{\bar q}}~
\int
{d{\bf k}_\perp \over 2(2\pi)^3}~
\Psi_\pi^{*}(x,{\bm
\kappa^\prime}_{\perp};\lambda^\prime_q,\lambda_{\bar q}) \nonu
\times~{\bar u(x,{\bf k}_\perp + {{\bm
\Delta}_\perp\over 2},\lambda^\prime_q)\over \sqrt{k^+ + {\Delta^+ \over 2}}}~
\gamma^+~ {u(x,{\bf k}_\perp - {{\bm
\Delta}_\perp\over 2},\lambda_q)\over \sqrt{k^+ - {\Delta^+ \over 2}}}
\Psi_\pi(x,{\bm \kappa}_{\perp};\lambda_q,\lambda_{\bar q})
\nonu=
\sum_{\lambda_q,\lambda_{\bar q}}~
\int
{d{\bm \kappa}_\perp \over 2(2\pi)^3}~
\Psi_\pi^{*}(x,{\bm
\kappa^\prime}_{\perp};\lambda_q,\lambda_{\bar q})~
\Psi_\pi(x,{\bm \kappa}_{\perp};\lambda_q,\lambda_{\bar q}),
\label{eq:overlap}
\ee
where  $u(x,{\bm \kappa}_{\perp},\lambda)$ is a LF Dirac spinor (see, e.g.
\cite{jaus}), and $\lambda_i$ are the   spin projections. The 
perpendicular component of the active quark momenta, ${\bf k}_\perp \pm {\bm
\Delta}_\perp/2$,  become
in the intrinsic frame  
\be
\label{eq:initialconf} 
{\bm \kappa}_{\perp }= {\bf k}_{\perp } - (1-x)~{{\bm \Delta}_\perp \over
2}, 
\qquad
{\bm \kappa^\prime}_{\perp } = {\bf k}_{\perp } + (1 -x)~{{\bm \Delta}_\perp \over
2}= {\bm \kappa}_{\perp }+(1 -x)~ {\bm \Delta}_\perp,
\ee 
with $x$ given by Eq. (\ref{kin}). Notice that  in the Drell-Yan frame $x_q=x$ (cf Eq.
(\ref{xb})), since $\xi=0$.

For the model calculation, we use a phenomenological LCWF which
satisfies Poincar\`e covariance and is eigenstate of the total angular
momentum operator in the Light-front dynamics.  As outlined in
Ref.~\cite{Chung:1988mu}, these properties can be fulfilled by
constructing the wave function as the product of a momentum wave
function $\psi(x,{\bm \kappa}_{\perp})$, which is spherically
symmetric and invariant under permutations, and a spin
wave function, which is uniquely determined by symmetry requirements.
Therefore, within LFHD one has 
\be
\label{eq:psifc}
 \Psi_\pi (x,{\bm \kappa}_{\perp };\lambda_q,\lambda_{\bar q}) 
= \psi_\pi(x,{\bm \kappa}_{\perp })~
\sum_{\mu_q,\mu_{\bar q}} \left(\oneh\mu_q\oneq \mu_{\bar q}|00\right)
{D}^{1/2\,*}_{\mu_q\lambda_q}\left[R_{M}({\bm \kappa})\right] 
{D}^{1/2\,*}_{\mu_{\bar q}\lambda_{\bar q}}\left[R_{M}(-{\bm \kappa})\right],
\label{pionwf}\ee
where ${\bm \kappa}\equiv \{{\bm \kappa}_\perp,\kappa_z\} $ with 
\be 
\kappa_z= M_0(x,{\bm \kappa}_\perp)~(x -{1 \over 2}),
\label{kappaz}\ee
and the free mass
defined by
\be M^2_0(x,{\bm \kappa}_\perp)= {m^2+|{\bm \kappa}_{\perp }|^2\over
x~(1-x)}~~~.
\label{kinelfhd}\ee  
The spin-dependent part contains the Melosh rotations 
$R_{M}({\bm \kappa})$ which convert the
  instant-form spins  of both  quark and antiquark 
into LF spins and ensure the rotational invariance of the pion
wave function.
The representation of the Melosh rotation is  explicitly given by
\be
{D}^{1/2}_{\lambda\mu}\left[R_{M}({\bm \kappa})\right ] =
\bra{\lambda}R_{M}({\bm \kappa})\ket{\mu} 
= \bra{\lambda}
\frac{m+xM_0(x,{\bm \kappa}_\perp)-i{\bm \sigma}\cdot(\hat{{\bf z}}\times{\bm \kappa}_\perp)}
{\sqrt{(m+xM_0(x,{\bm \kappa}_\perp))^2+{\bm \kappa}_\perp^2}}\ket{\mu}.
\ee

For the momentum-dependent part of the pion wave function we adopt
the following exponential form used in Refs.~\cite{Chung:1988mu,Schlumpf:1994bc} 
\be
\label{eq:psifc2}
 \psi_\pi(x,{\bm \kappa}_{\perp })  
=[2(2\pi)^3]^{1/2}
\left(\frac{M_0(x,{\bm \kappa}_\perp)}
{4~x (1-x)}\right)^{1/2}\frac{1}{\pi^{3/4}\beta^{3/2}}
\exp{(-\kappa^2/(2\beta^2))}.
\ee
The wave function in Eq.~(\ref{eq:psifc2}) is normalized as $$\int_0^1~dx
\int {d{\bm \kappa}_\perp\over 2(2\pi)^3}~|\psi_\pi(x,{\bm \kappa}_{\perp })|^2=1$$ (reminding
that $d\kappa_z= dx~ M_0(x,{\bm \kappa}_\perp)/[4 x (1-x)]$), and 
 depends on the free parameter $\beta$ and the quark mass $m$, which have been
fitted to the pion charge radius and decay constant.

Inserting the model wave function of Eq.~(\ref{pionwf}) in the LCWF overlap 
representation of GPD in Eq.~(\ref{eq:overlap}), one obtains
\be
H^u(x,\xi=0,t)  = 
\int
{d{\bm \kappa}_\perp \over 2(2\pi)^3}~
\psi_\pi(x,{\bm \kappa^\prime}_{\perp})
\psi_\pi(x,{\bm \kappa}_{\perp})~
{m^2 +{\bm \kappa}^\prime_{\perp} \cdot {\bm \kappa}_{\perp} \over x~(1-x)~ M_0(x,{\bm
\kappa}^\prime_\perp)~M_0(x,{\bm \kappa}_\perp)}\ . \nonu
\label{lfhdgpd}
\ee

In the forward limit $\Delta^\mu\rightarrow 0$, the Melosh rotation matrices
combine to the identity matrix and one obtains the ordinary parton 
distribution as momentum density distribution given by the square of 
the momentum-dependent part of the wave function \cite{brodsky}, i.e. for $
x\geq 0$ one gets
\be
u(x) =
\int~
{d{\bm \kappa}_\perp\over 2(2\pi)^3}~|\psi(x,{\bm \kappa}_{\perp})|^2\ .
\ee

\section{Results and Discussion}
\label{Ris}
In this Section the results obtained from the different models described in the
previous Sections are presented and discussed.
Let us first illustrate the actual values of the
parameters entering the  three models. 

For the
covariant model (Sec. \ref{SYM})
the CQ mass and the pion mass have  values $m=220$ MeV and $m_\pi=140$ MeV, 
respectively. It should be pointed out that, for some runs, the value $m_\pi=0$
 has been used  in order to match the vanishing pion mass adopted for the
Mandelstam-inspired model (see Sec. \ref{LFMan}). This change will be adequately
emphasized whenever applied (in this case the CQ mass is a little bit lowered,
 i.e. $m=210$ MeV). The parameter $m_R$ present in 
  the pion Bethe-Salpeter
amplitudes is fixed through the pion decay constant, obtaining  $m_R=600$ MeV
for the  sum-form (Eq. (\ref{vertexs}))  and $m_R=1200 $ MeV for the
product-form (Eq. 
(\ref{vertexp})).  

In the Mandelstam-inspired model, as already mentioned, all the parameters 
are the same ones used in
\cite{DFPS}, except for i) $w_{VM}=-1$ that yields the relative weight of the
instantaneous contributions and ii) the CQ
mass, $m=200$ MeV, i.e.   the one
adopted in \cite{nuc08} within the same approach for the very detailed description of  the nucleon 
em form factors in both the spacelike and 
timelike region. As already mentioned, given the
complexity of the calculation a simplifying assumption of
a vanishing pion mass has been also added. Finally, in the VM dressing 
of the
quark-photon vertex (cf Eq. (\ref{VMD})) up to 20 isovector mesons have been 
considered in order to 
have a good
convergence even for $t=-10$ (GeV/c)$^2$. 

In the LFHD model (see
 Sec. \ref{LFHD}), a CQ mass  $m=250$ MeV and a
wave-function parameter $\beta=319.4$ MeV have been used in order to 
reproduce the pion charge
radius ($r_{ch}=0.670\pm0.02~fm$) and the pion decay constant~
\cite{Schlumpf:1994bc}.

\begin{figure}[thb]

\vspace{1cm}
\includegraphics[width=8.5cm]{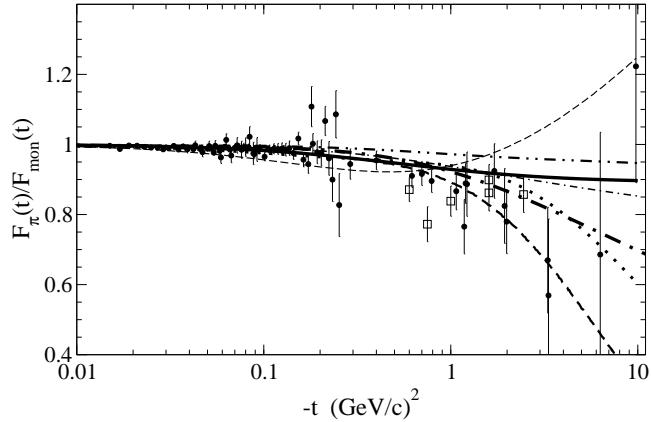}
\caption{ Pion form factor vs $-t$. 
Thin dashed line: covariant symmetric model of Ref. \cite{pach02}, with the momentum
dependence of the pion Bethe-Salpeter amplitude given by the sum-form  of Eq.
 (\ref{vertexs}),
and   $m_\pi=140$ MeV. Double-dot-dashed line: calculation performed within 
the LF Mandelstam-inspired model (cf Sec. \ref{LFMan}),  
 by using  an asymptotic pion wave function \cite{lepage} 
 with $m_\pi=0$, and adopting a CQ mass of $m=200$ MeV (notice that in  
  \cite{DFPS} $m=265 $ MeV). Thick solid line:  monopole fit to Lattice data  as obtained 
  in Ref. \cite{brom07}, arbitrarily extended in this figure
from $-4$ (GeV/c)$^2$ to $-10$ (GeV/c)$^2$ (see text). 
Thick  dot-dashed  line:  faster-than-monopole fit to Lattice data  as obtained 
  in Ref. \cite{brom07}, arbitrarily extended in this figure
from $-4$ (GeV/c)$^2$ to $-10$ (GeV/c)$^2$ (see text).
Dot-dashed line: the same as the double-dot-dashed line, but with a non
perturbative pion wave function, eigenstate of the squared LF mass operator of Ref.
\cite{FPZ}.  Dotted line: the same
as the thin dashed line, but with the
product-form of Eq. (\ref{vertexp}) for  the pion Bethe-Salpeter 
amplitude. Thick dashed line: LFHD model (cf Sec. \ref{LFHD}) with a Gaussian
pion wave function and the proper Melosh rotations. Experimental data: 
full dots from the
collection of Ref. \cite{baldini}; open squares,  TJLAB data from Ref.
\cite{TJLAB}.}
\label{ffpfig} 
\end{figure}
\begin{figure}
\includegraphics[width=8.5cm]{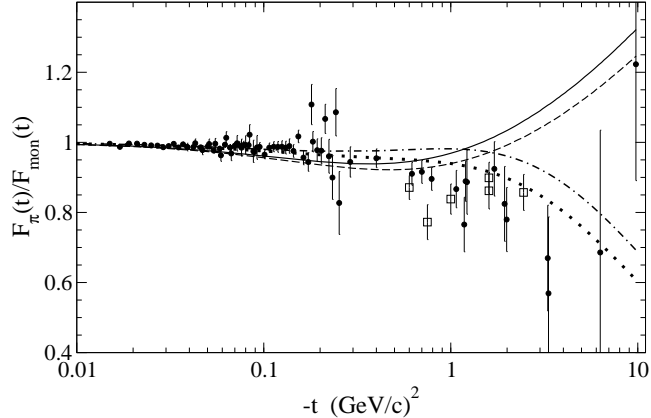}
\caption{Pion form factor calculated  within the covariant model of Sec.
\ref{SYM}, with and without a vanishing pion
mass. Solid line: sum-form for  the 
pion Bethe-Salpeter amplitude, Eq. 
(\ref{vertexs}), and $m_\pi=0$. Dashed line: the same as the solid
line, but with $m_\pi=140$ MeV. Dot-dashed line: product-form for  the  
pion Bethe-Salpeter amplitude, Eq. 
(\ref{vertexp}), and $m_\pi=0$. Dotted line:
the same as the dash-dotted
line, but with $m_\pi=140$ MeV. Experimental data as in Fig. \ref{ffpfig}.}
\label{ffpfig2} 
\end{figure}
\begin{figure}
\includegraphics[width=8.5cm]{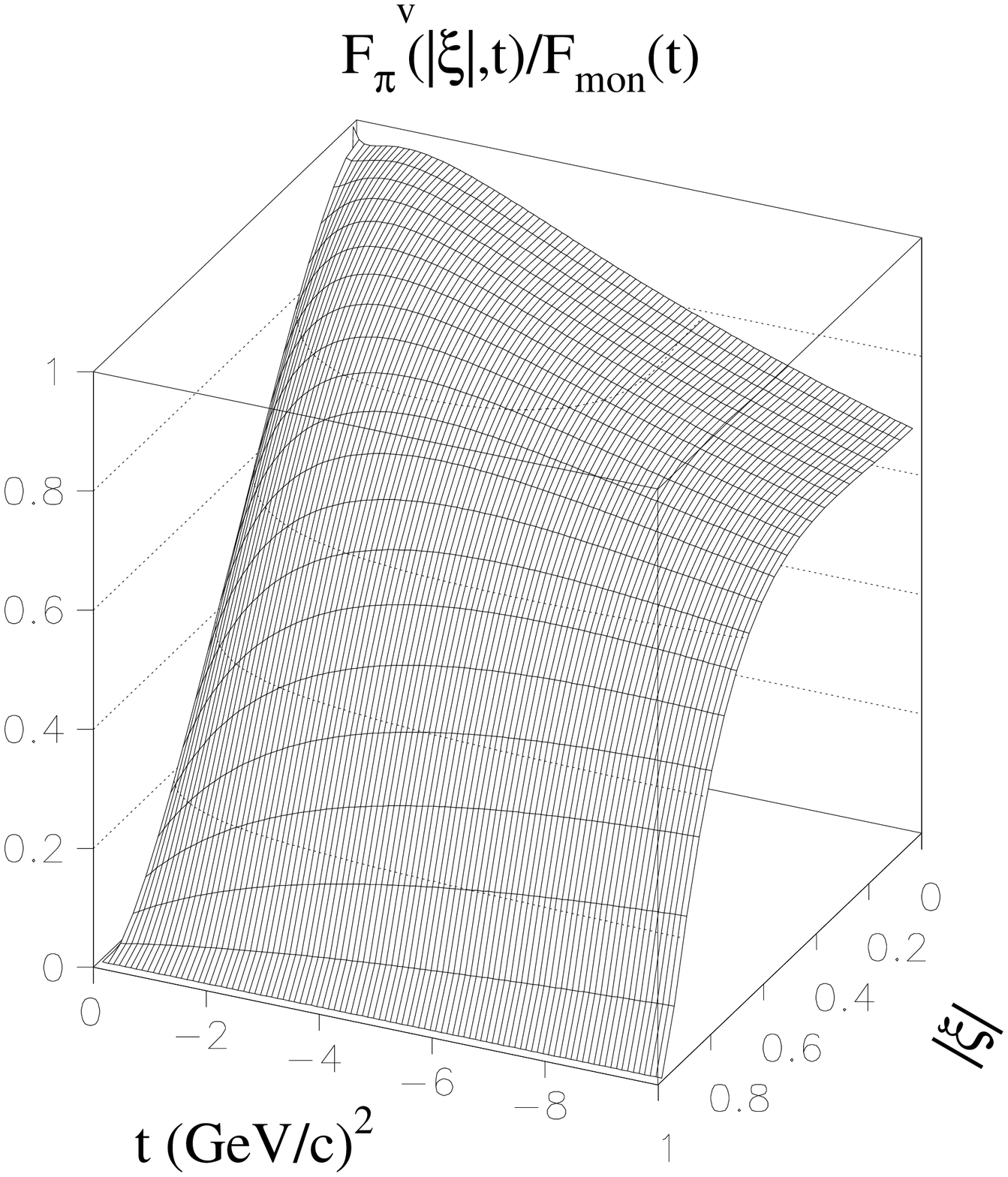}~~~
\includegraphics[width=8.5cm]{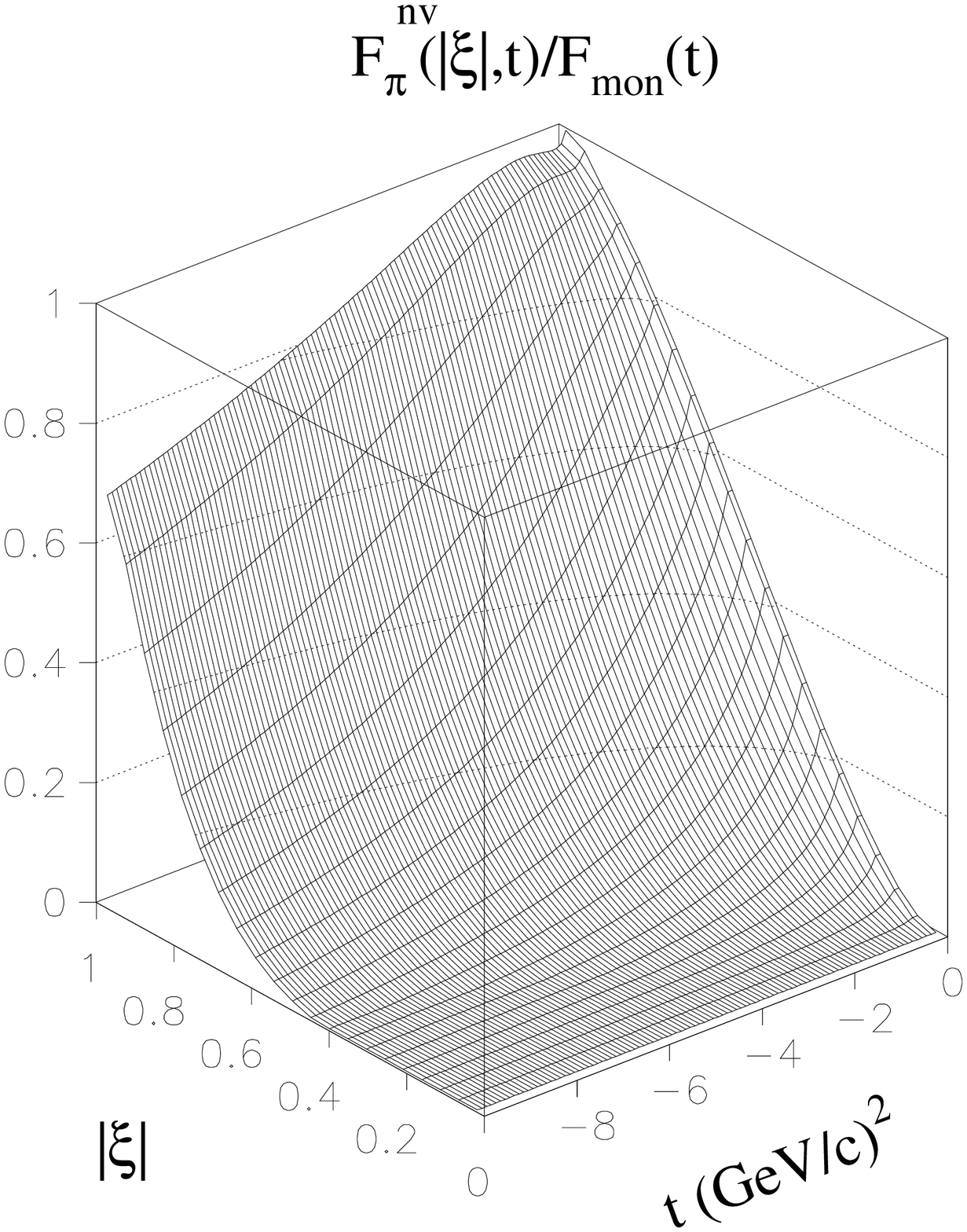}
\caption{Left Panel:  valence contribution, $F^{(v)}_\pi (|\xi|,t)$, 
 to the em pion form factor (see 
Eq. (\ref{srval})), evaluated within the covariant 
symmetric model of Sec. \ref{SYM} by using the product-form for the 
momentum-dependent
part of
the Bethe-Salpeter
amplitude (cf  Eq. (\ref{vertexp})) and choosing $m_\pi=0$ for covering the
whole range $0\leq |\xi|\leq 1$, according to Eq. (\ref{nompi}).
Right Panel: the same as in the
Left Panel, but for the non-valence
contribution, $F^{(nv)}_\pi (|\xi|,t)$, (see Eq. (\ref{srnval})). Note the 
different
orientations of the axes in the two Panels, for a straightforward selection 
of  the
relevant regions.}
\label{ffpixi} 
\end{figure}
First of all, 
the theoretical models have been compared with   available experimental data,
in particular 
 the pion em
form factor in the spacelike region.

In Fig.
\ref{ffpfig}, it is shown the ratio between the spacelike  form factors,
calculated by using our models, and
the monopole
form factor $F_{mon}=1 /(1+|t|/m_\rho^2)$ ($m_\rho=.770$ GeV).  The
relevance of  such a  presentation of the form factor is twofold:
i) dividing by $F_{mon}$ allows one to   avoid the log plot 
that
hinders a detailed analysis, ii) more important, one can immediately 
discriminate 
between models that produce a  divergent charge density at short distances 
and models that do not (cf, e.g., \cite{miller09}), since their fall-off is more
rapid than $F_{mon}$. 

For the sake of completeness, we have also displayed two different fits 
(thick solid and dot-dashed lines
  in Fig. \ref{ffpfig}) to  the Lattice data as
obtained in Ref. \cite{brom07}. In that paper, Lattice data
 have been 
extrapolated to the experimental pion mass, and  they
were
 described up to $t=-4$ (GeV/c)$^2$ both in terms of i) a monopole function
 $F^{lat}_\pi(t)=1/[1-t/M^2(m^{phys}_\pi)]$ with $M(m^{phys}_\pi)=0.727 $ GeV
 and ii) a function with a fall-off faster than the monopole one, i.e. 
$F^{lat}_\pi(t)=1/[1-t/(p~M^2(m^{phys}_\pi)]^{p}$ 
with $p=1.173\pm 0.058$ and $M(m^{phys}_\pi)=0.757\pm 0.018 $ GeV.
  In Fig.
\ref{ffpfig} the Lattice results have been arbitrarily extended by using the
previous functions  
from $t=-4 $ (GeV/c)$^2$ to $t=-10$ (GeV/c)$^2$, 
with a quite reasonable outcome.

To show the sensitivity 
of the covariant model of Sec. \ref{SYM} upon the change of the
pion mass, a comparison between
calculations performed with a
vanishing pion mass and with $m_\pi=140$ MeV is presented in Fig. \ref{ffpfig2}. 
These calculations  are helpful in view
of the  following comparisons with the Mandelstam-inspired model, 
where the value $m_\pi=0$
has been adopted.  It is interesting to notice from Figs. \ref{ffpfig} and 
 \ref{ffpfig2} that the sum-form for the BS amplitude is unable to accurately 
 describe the experimental 
 em form factor at high values of $|t|$.

In order to illustrate the frame dependence of the Fock decomposition of the 
em
form factor, in Fig. \ref{ffpixi} the valence and non-valence contributions to
the pion form factor (Eq. (\ref{srtot})) within the covariant model based 
on the product-form and
$m_\pi=0$ are presented. Such a choice for $m_\pi$ is suggested (cf Eq.
(\ref{nompi})) by the 
need  to explore the whole range  
$0\leq |\xi| \leq 1$. The sum of the two contributions becomes $\xi$
independent, and the result is shown in  Fig. \ref{ffpfig2} by the dot-dashed 
line. Figure \ref{ffpixi} allows  us to disentangle the valence and non-valence contributions.  Indeed, different values of $\xi$ correspond to
  different choices of the frame (let us remind that $\xi=0$ corresponds to the
  Drell-Yan frame and $\xi=-1$ to the frame where ${\bm \Delta}_\perp=0$).
  Moreover, it is worth noting that  the operator "number of constituents" does
  not commute with the whole set  of the  Poincar\'e generators, and therefore
 a change of  frame alters the non-valence content. 
The knowledge of valence and non-valence  contributions in the plane $(\xi,t)$ could impose 
new constraints to  models that aim to go
beyond the  standard CQM.

\begin{figure}
\includegraphics[width=8.5cm]{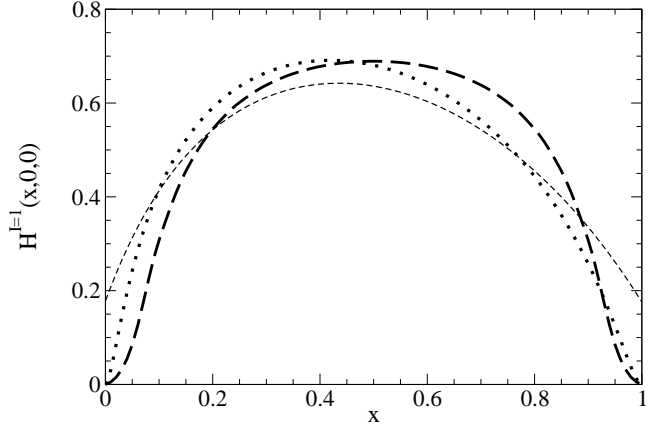}
\caption{ Isovector GPD $H^{I=1}(x,0,0)$, equal to half of the parton
distribution (see Eq. (\ref{pdf})), vs $x$. Thin dashed line:
covariant model of Sec. \ref{SYM}, calculated by using the sum-form, 
Eq. (\ref{vertexs}), for  the 
pion Bethe-Salpeter amplitude and $m_\pi=140$ MeV. Dotted line: the 
same as the
thin dashed line, but for the product form, Eq. (\ref{vertexp}). 
Thick dashed line: LFHD model 
of Sec. \ref{LFHD}, with a Gaussian pion wave function and the proper Melosh
rotations.
 The variable $x$, given in Eq. (\ref{kin}), coincides with the usual
LF longitudinal fraction $x_q$, since $\xi=0$ (see text  below Eq. (\ref{struc})).}
\label{strucx} 
\end{figure}

After completing the analysis of the em form factor within our models, 
in Fig. \ref{strucx}, the isovector GPD for positive $x$, namely 
$H^{I=1}(x,0,0)=u_v(x)/2$ 
(see Eq. (\ref{pdf})), is shown as a function of $x\equiv x_q$ (since for 
$\xi=0$ one recovers the longitudinal momentum fraction, Eq. (\ref{xb})).
 It should be pointed out that, at this stage of our analysis, no evolution
  has
been applied.  The effects of the evolution for the parton distribution  
will be considered elsewhere,
together  with  a study  of the evolution for the whole GPD. Calculations 
for the
covariant model of Sec. \ref{SYM} and the LFHD model of Sec. \ref{LFHD} are 
shown in Fig. \ref{strucx}. Notice that  the Mandelstam-inspired LF model presently allows
predictions only for $|\xi|=1$. In order to extend to $\xi=0$ this approach, a
 non-vanishing  value of $m_\pi$  and a bare term,
 besides the VMD one, should be considered. Thus one can  take
into account the contribution depicted in Fig. \ref{figbs}(a), that 
produces the
valence term in $H^{I=1}(x,\xi,t)$, but   
   new, non-trivial parameters have to be added (cf the nucleon case in
   \cite{nuc08}).

The comparison  in  Fig. \ref{strucx} shows the  
difficulty  of the sum-form, Eq. (\ref{vertexs}), for 
the pion Bethe-Salpeter amplitude to give a
realistic parton distribution, i.e. to have a vanishing value at the end-points.
Reminding that, for $\xi=0$, the presence of the delta-function in Eq. (\ref{jmu}) and the kinematical 
relations in
Eq.  (\ref{kin}) impose the correct support $[0,1]$ for the variable $x$ (remind
that for $\xi=0$, one has $x=x_q$), the
sum-form produces a discontinuity at the end-points, i.e. an infinite
derivative. It is instructive to correlate such a drawback to the one already seen 
in Fig. \ref{ffpfig},
where the sum-form is not able to reproduce  the em form factor  at    
high values of $(-t)$. Indeed, in both cases, the high momentum part of the valence
component of the pion state is involved. As a matter of fact, for $x=0$ and 
$x=1$, the intrinsic
three-momentum becomes infinite (cf Eqs. (\ref{kappaz}) and  (\ref{kinelfhd})), and therefore  
small distances are involved, just as in the case  of  the tail of the
em form factor, where  the influence upon  the small-$r$ part of the pion wave
 function is felt. The
more realistic behavior of the product-form (\ref{vertexp}), can be ascribed 
to
a $|{\bf k}_\perp|$ fall-off  like the one dictated
 by a BS kernel dominated by a
one-gluon-exchange, as already pointed out in Sec.
\ref{SYM}. An important,  final remark is the clear shift towards small $x$ 
of the
curves evaluated within the covariant model, while   the prediction obtained
within the LFHD model is symmetric with respect to $x=1/2$. 
Such an interesting difference could be explained by the fact that 
the full covariance of the
model of Sec. \ref{SYM} together with its dynamical content, related to the 
adjusted parameter $m_R$,  could take into account some effects
 beyond the pure $q\bar q$ component of the pion state. First, one should note
 that  the
 valence component, Eq. (\ref{phival}),
 generates a quark distribution symmetric with respect to $x=1/2$ and a
 probability definitely less than 1: for the sum-form $P_{val}=0.78$ and  
 for the product-form $P_{val}=0.84$ (see also \cite{pach02}). Then, by using the Fock decomposition of the pion
 state (see, e.g., \cite{diehlpr} for a general discussion), one immediately 
 recognizes contributions from both the $q\bar q$ component,
 (i.e. the  valence
 component) and from other components with more constituents 
 (see e.g. the instantaneous contributions in  Eqs. (\ref{i1}), (\ref{i2}) 
 and (\ref{i3}) and the analysis in \cite{adnei}). 
 Thus, the active quark shares
 the longitudinal momentum of the pion  with more than one spectator parton, 
 belonging to the Fock space configuration
beyond the valence one. Therefore, the shift    toward values of $x$ less than
  $1/2$ is expected, since  our covariant model contains more physical
  effects than the basic one.
 In particular, for a non vanishing pion mass the average longitudinal 
 momentum
fraction for the sum-form is  $<x_q>\sim
0.483$ and for the product-form is  $<x_q>\sim
0.471$, i.e. quite similar, but a little bit different from $1/2$. As a
simple cross-check we have   reobtained those values also from 
$A^{I=0}_{2,0}(0)=<x_q>$
 (cf  Eq. (\ref{oddm}) with $j=0$ and
Eq.
(\ref{struc1})). 

\begin{figure}
\includegraphics[width=8.5cm]{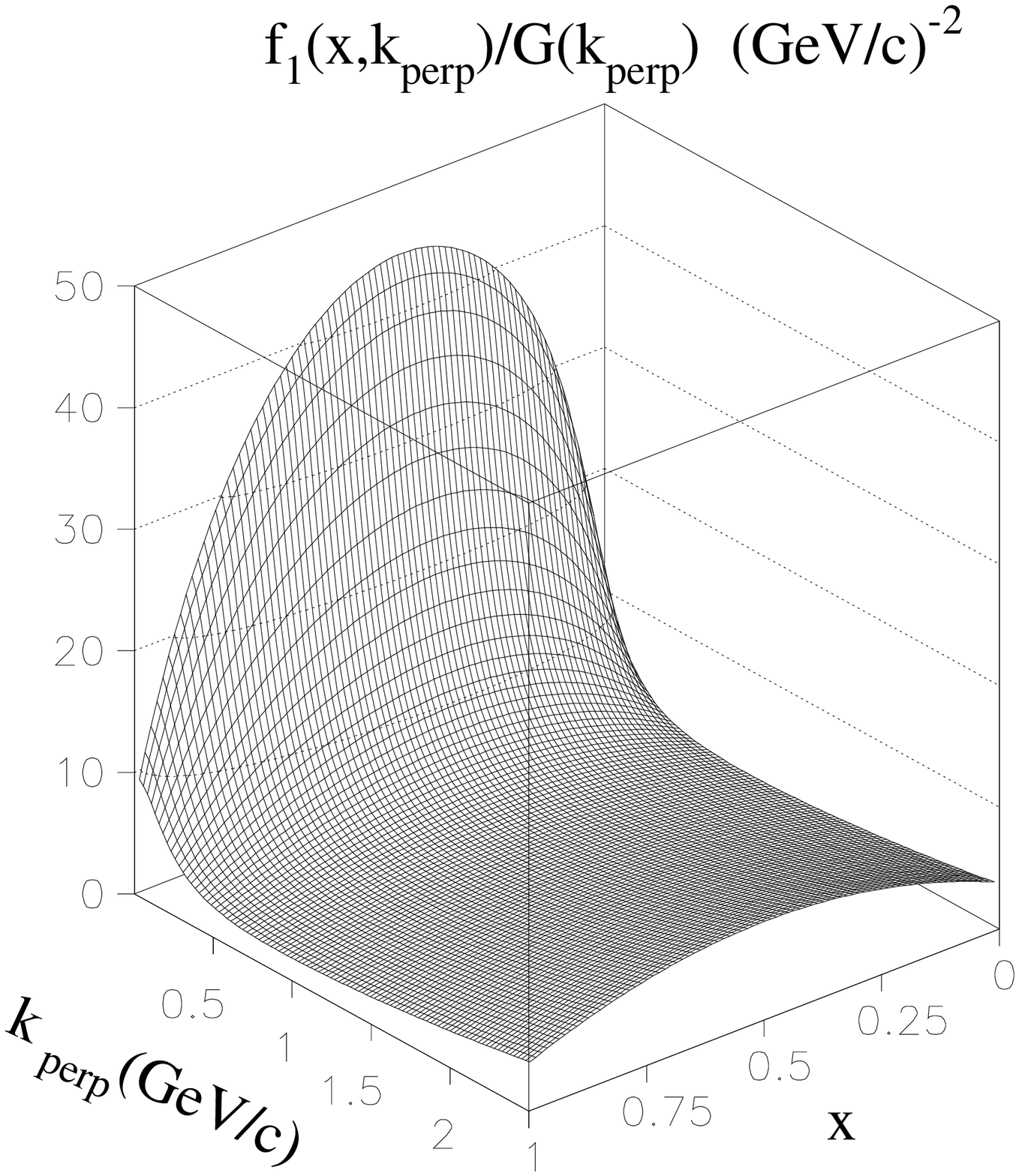}~~~
\includegraphics[width=8.5cm]{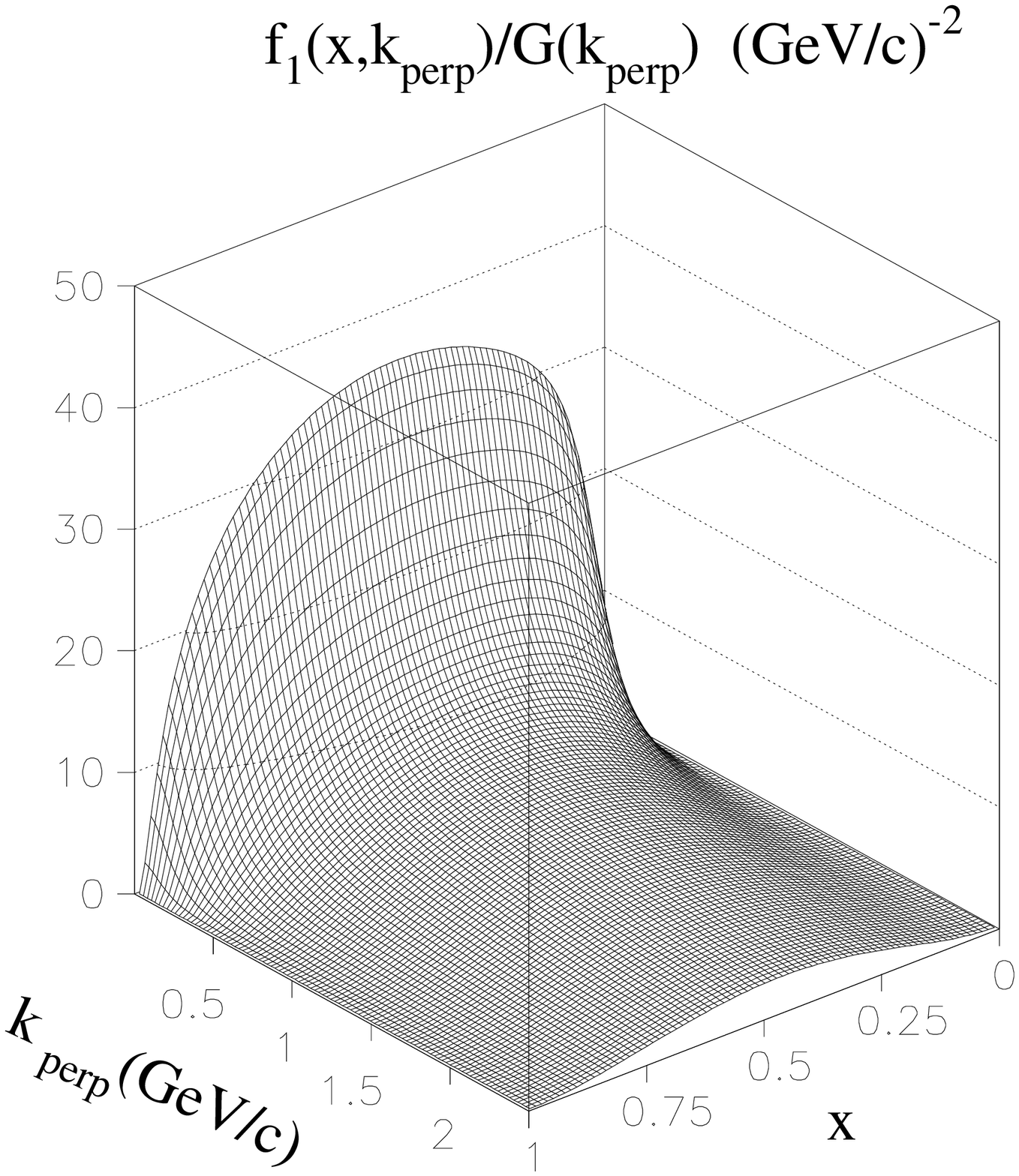}
\caption{Transverse-momentum dependent function, 
$f_1(x,|\bm{k}_\perp|^2)/G(|\bm{k}_\perp|)$, with $G(|\bm{k}_\perp|)
=1/ (1+|\bm{k}_\perp|^2/m^2_\rho)^4$.
Left Panel:  sum-form   of  the Bethe-Salpeter
amplitude (see
Eq. (\ref{vertexs})). Right Panel: the same as in the
Left Panel, but for  the product-form   of  the Bethe-Salpeter
amplitude (see
Eq. (\ref{vertexp})). The normalization is given by 
$\int_0^1dx\int d{\bf k}_\perp~ f_1(x,|{\bf k}_\perp|^2)=1$, and $k_{perp}$ means 
$|{\bf k}_\perp|$.}
\label{f1xkfig} 
\end{figure}

A more detailed analysis of the parton distribution can be achieved by using
the chiral-even TMD distribution, $f_1(x,|{\bf k}_\perp|)$, see Eq. 
(\ref{f1xk}).
 In Fig. \ref{f1xkfig}, the TMD distributions calculated within the 
 covariant
 model by using the different BS amplitudes of Eqs. (\ref{vertexs}) and 
(\ref{vertexp}) are shown. In order to avoid log plot, 
$f_1(x,|{\bf k}_\perp|)$
has been divided by $G(|\bm{k}_\perp|)
=1/ (1+|\bm{k}_\perp|^2/m_\rho^2)^4$, (with $m_\rho=770$ MeV)). Clearly, 
the product-form has a
$|{\bf k}_\perp|$ fall-off faster than the sum-form does, i.e. low
transverse-momentum partons are favored in the first case. 

\begin{figure}
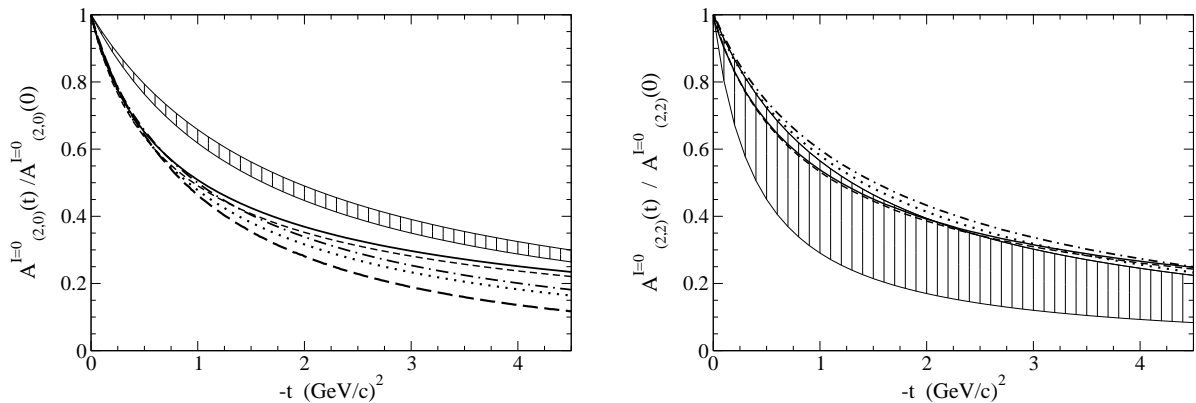

\includegraphics[width=7.5cm]{Fig8a.eps} ~~~~
\includegraphics[width=7.5cm]{Fig8b.eps}
\caption{ Left Panel: the ratio   $A^{I=0}_{2,0}(t)/A^{I=0}_{2,0}(0)$, 
involving
the generalized form factor $A^{I=0}_{2,0}(t)$ that appears in the second moment of the 
isovector  GPD $H^{I=0}$ (cf
Eqs. (\ref{gravff}) and (\ref{oddm}))  as a
function of $t$. Solid line: sum-form for  the 
pion Bethe-Salpeter amplitude, Eq. 
(\ref{vertexs}), and $m_\pi=0$. Dashed line: the same as the solid
line, but with $m_\pi=140$ MeV. Dot-dashed line: product-form for  the  
pion Bethe-Salpeter amplitude, Eq. 
(\ref{vertexp}), and $m_\pi=0$. Dotted line:
the same as the dash-dotted
line, but with $m_\pi=140$ MeV. Thick long-dashed line: LFHD model (cf Sec. \ref{LFHD}) with a Gaussian
pion wave function and the proper Melosh rotations.
 Shaded area:  results
from Lattice QCD \cite{thesislattice08}(see text).  Right Panel: the same as the Left Panel,
 but for  $A^{I=0}_{2,2}(t)/A^{I=0}_{2,2}(0)$. }
\label{figsecm} 
\end{figure}

The analysis of both the generalized form factors involved in the  
second moment 
of the isoscalar pion GPD, i.e. $A^{I=0}_{2,0}(t)$ and  $A^{I=0}_{2,2}(t)$ 
(cf 
Eq. (\ref{oddm}) with $j=0$), has to
be performed necessarily  within the covariant analytic model of 
Sec. \ref{SYM}.
This is obvious if we look at Eq. (\ref{gravff}), where the polinomiality
imposes a square dependence upon $\xi$, and therefore one needs a model that
covers an extended  range for the variable $\xi$. Indeed, 
for each value of $t$, we have first
numerically checked the parabolic behavior against $\xi$, and then we have
extracted the 
  coefficients  of the parabolic fit  getting the values of 
$A^{I=0}_{2,0}(t)$ and  $A^{I=0}_{2,2}(t)$. 
  Figure \ref{figsecm} shows a
comparison between i) recent results from Lattice QCD, 
extrapolated to the physical pion mass
\cite{thesislattice08,brommelprl},  ii) our covariant calculations evaluated with both $m_\pi=0$ and  
$m_\pi=m_{phys}$ by using  
the sum- and the product-form for
the BS amplitude (Eqs. (\ref{vertexs}) and (\ref{vertexp})) and iii)  the LFHD 
result   (see
 Sec. \ref{LFHD}) for $A^{I=0}_{2,0}(t)$ only, since  this approach at the
 present stage
 allows one to perform calculations exclusively for $\xi=0$. Indeed,
the ratios $A^{I=0}_{2,0}(t)/A^{I=0}_{2,0}(0)$ 
and $A^{I=0}_{2,2}(t)/A^{I=0}_{2,2}(0)$ are presented  in order to get rid of 
the evolution (see  Ref. \cite{bronio08} for a detailed discussion of this 
issue).  
 The Lattice calculations are 
described through
a monopole form, $1 /(1-t/M^2_{2,i})$, as obtained in \cite{thesislattice08} 
from the  analysis of  their Lattice data, 
 without evolution and with evolution 
in the
$\overline{MS}$ scheme at the scale $\mu=2$ GeV. In particular,  we have used the following
values:
$M_{2,0}=1.329\pm 0.058$ GeV and  $M_{2,2}=0.89\pm 0.25$ GeV, corresponding to
an analysis of the Lattice data that satisfies  the low-energy theorem, i.e.
$A^{I=0}_{2,0}(0)=-4A^{I=0}_{2,2}(0)$. The
uncertainties on the previous masses generate the shaded areas  in the 
Left and Right panels in Fig. \ref{figsecm}. 
 
Unfortunately, i) the available range of $(-t)$ (we refrained to enlarge the
interval as we did in the case
of the em form factor, since we do not have experimental data yielding
confidence in an arbitrary extension of the monopole fit) and ii) the large
 uncertainties in the
Lattice calculations of $A^{I=0}_{2,2}$ do not allow us to elaborate too much on
the comparison between our phenomenological models and the Lattice results. 
On the
other hand, for large values of $|t|$ the calculations obtained by using the
covariant model with the product-form and $m_\pi=140$ MeV could give some
insight on the expected behavior of the Lattice calculations, since 
 one could argue that the covariant model with the product-form
phenomenologically contains at some extent dynamical features typical of  QCD,
like the one-gluon-exchange dominance at small distances. 
In order to complete the 
information, in
Table I  the values of $A^{I=0}_{2,0}(0)$ and 
$A^{I=0}_{2,2}(0)$ are shown. It is worth noting that while the
Lattice calculations largely fulfill  the low-energy theorem, as already
mentioned,  our calculations do not.
Furthermore, it should be pointed out that for small $t$ the
disagreement between Lattice data and the calculation with the covariant 
approach at
some extent is an expected one, since the mechanism responsible for the
confinement is not present in our model, and therefore we have a free 
propagation of the
$q\bar q$ pair. A possible solution could be elaborated following the 
suggestion
in Ref. \cite{silviavm}, where a covariant model without the disturbing 
free propagation of 
the  $q\bar q$ pair was proposed and applied to the em decays of the vector
mesons.
\begin{table}
\label{grff}
\caption{Gravitational form factors at $t=0$, (cf Eq. (\ref{oddm}) with $j=0$), obtained i) 
within the covariant
model of Sec. \ref{SYM} and both the sum- and the product-form for
the BS amplitude, Eqs. (\ref{vertexs}) and (\ref{vertexp}); and  ii) from the Lattice data of Ref.
 \cite{thesislattice08}.}
 \begin{center}
\begin{tabular}{|c||c|c|c|c|c|c|}
\hline
~ & Sum & Sum  & 
Product & Product  &
Latt. no evol.  & Latt. with $\overline{MS}$ evol.    \\
&  $m_\pi=0$  &  $m_\pi=m_{phys}$  & 
 $m_\pi=0$ &  $m_\pi=m_{phys}$ &
 $m_\pi=m_{phys}$ & $m_\pi=m_{phys}$\\
\hline  
$ A^{I=0}_{2,0}(0)$   & 0.4828 &  0.4833 & 0.4707 & 0.4710 & 0.365 & 0.261\\
\hline
$A^{I=0}_{2,2}(0)$    & -0.0307 & -0.0272   & -0.0357 & -0.0327 & -0.092 
& -0.066 \\
\hline 
\end{tabular}
\end{center}
\end{table}

\begin{figure}[t]
\includegraphics[width=8.0cm]{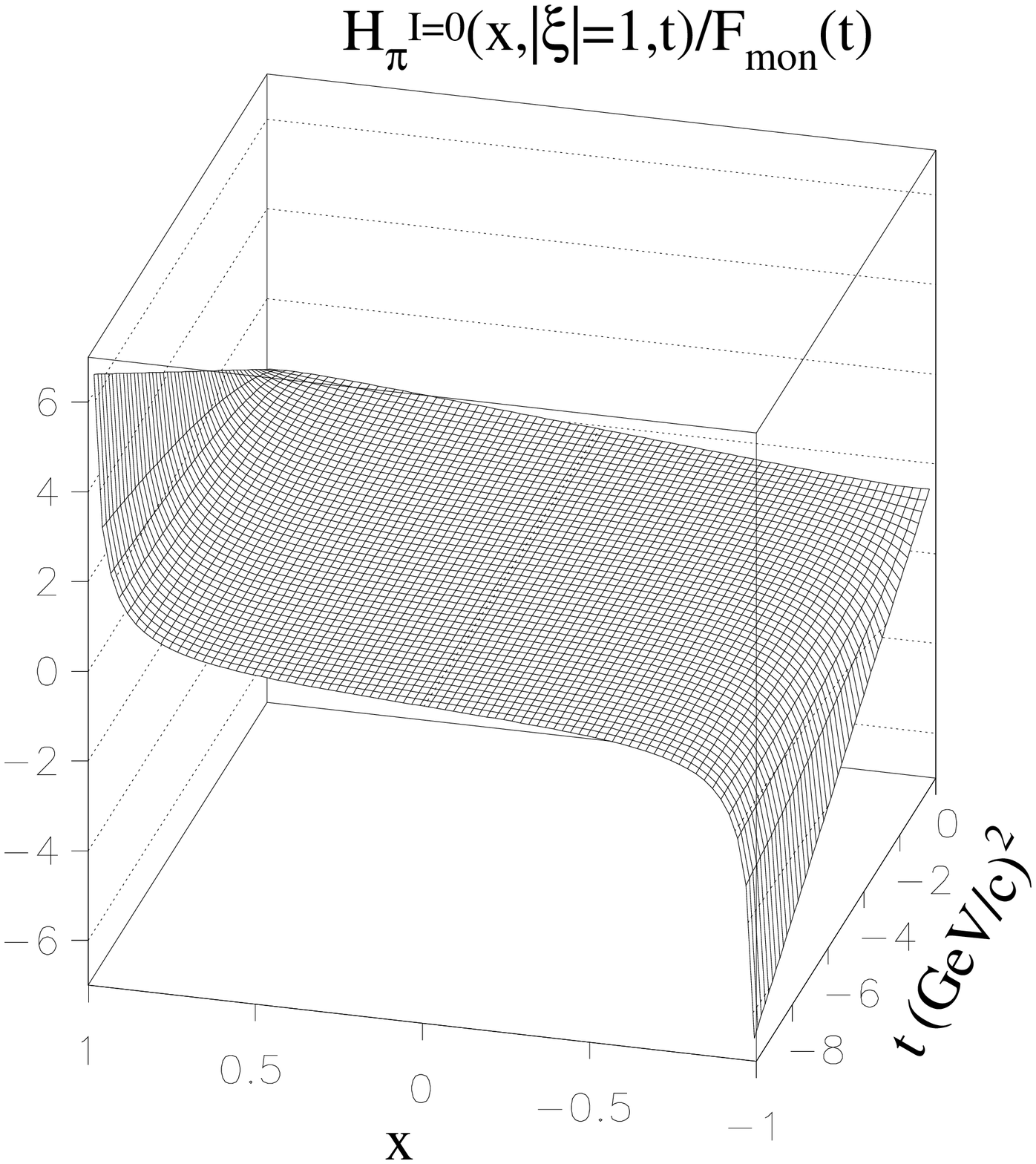}~~~
\includegraphics[width=8.0cm]{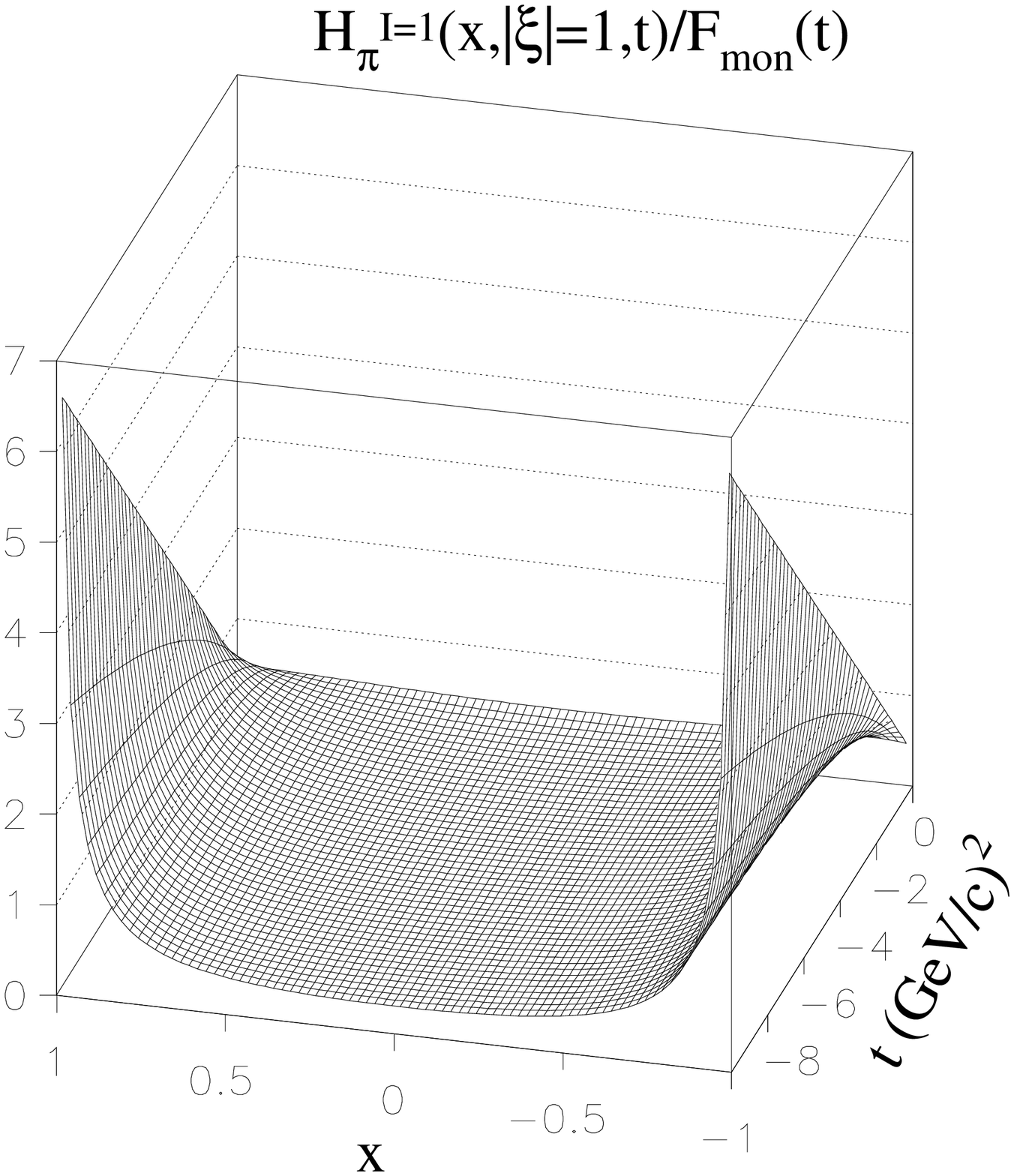}

\vspace{-3cm}
\includegraphics[width=8.0cm]{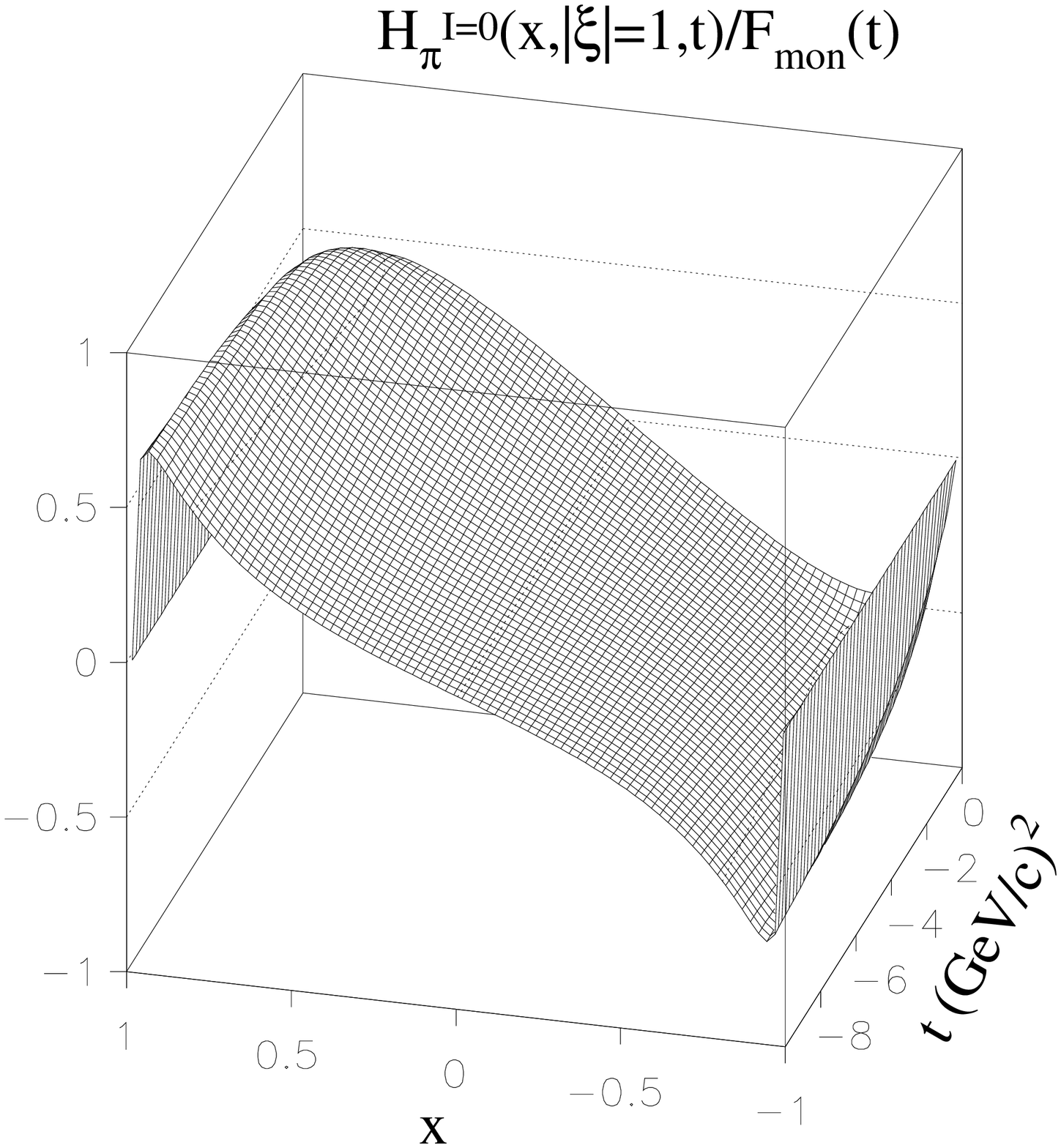}~~~
\includegraphics[width=8.0cm]{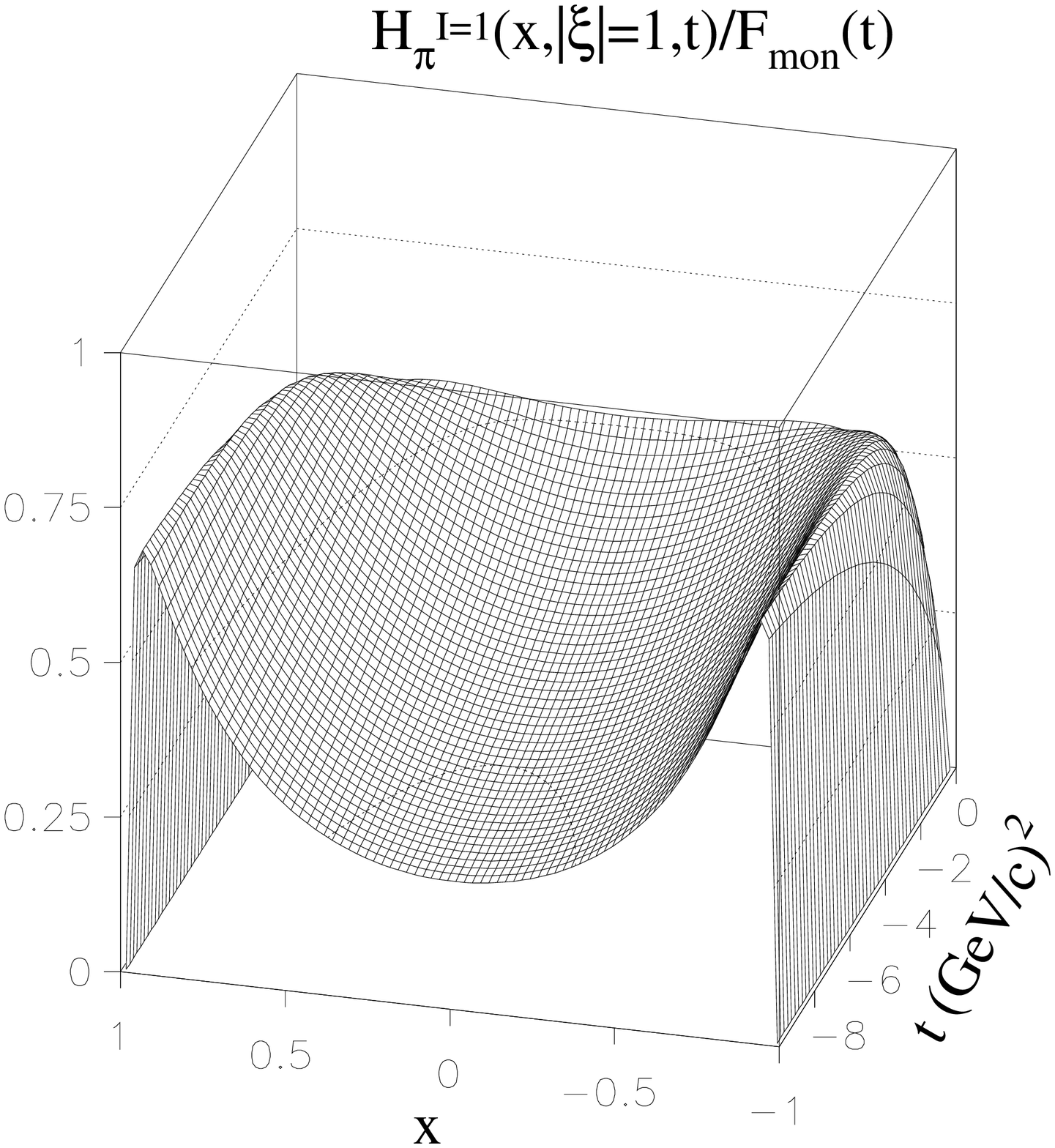}
\caption{Upper Left Panel: Isoscalar no-helicity flip GPD from the covariant 
symmetric model of Sec. \ref{SYM} with the sum-form for the Bethe-Salpeter
amplitude (Eq. (\ref{vertexs})) at
$|\xi|=1$. The value of $\xi$ is fixed by using 
$m_\pi=0$ (cf Eq. (\ref{nompi})), for the sake of comparison with the
microscopic 
model of Sec. \ref{LFM}, whose results are shown in Fig. \ref{Hvmfig}. On the
z-axis the ratio  with respect to $F_{mon}=1 /(1+|t|/m_\rho^2)$ is presented.
 Upper Right Panel:
the same as in the Upper
Left Panel, but for the isovector GPD.
Lower Panels: the same as in Upper Panels, but for the product-form for 
the Bethe-Salpeter
amplitude (see
Eq. (\ref{vertexp})).}

\label{Hsymsfig} 
\end{figure}

\begin{figure}
\includegraphics[width=8.0cm]{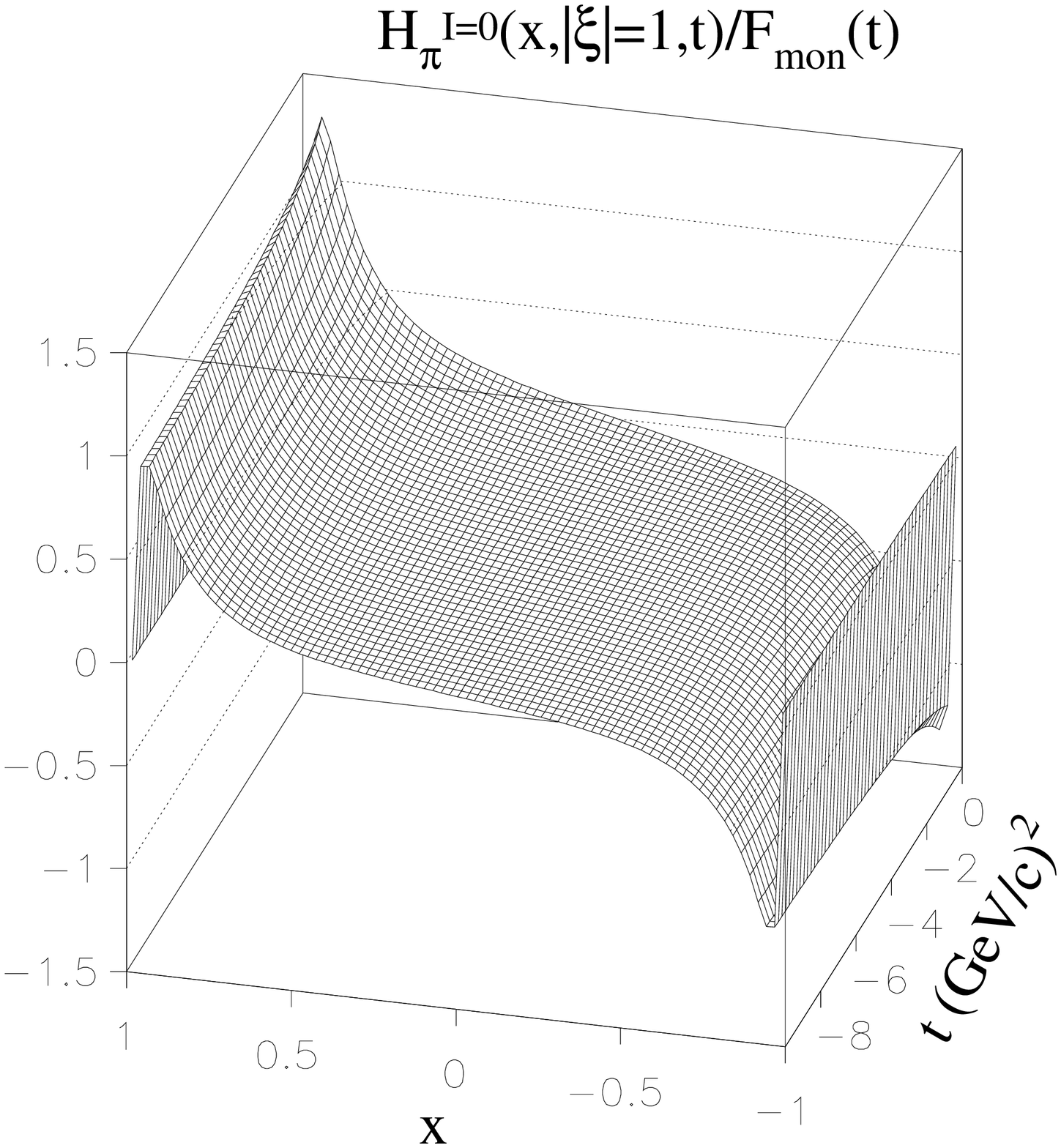}~~~
\includegraphics[width=8.0cm]{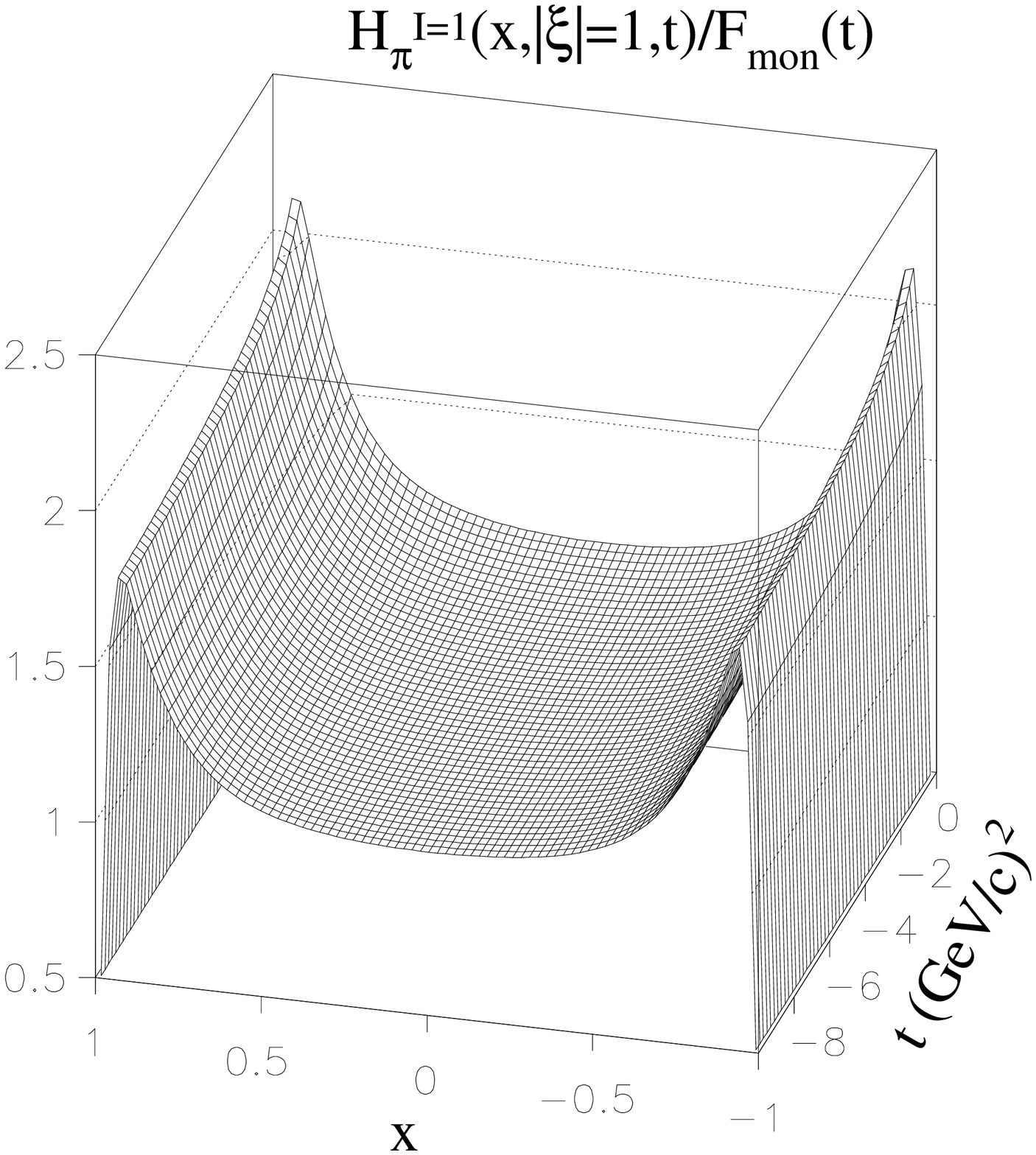}
\caption{Left Panel: Isoscalar no-helicity flip GPD from the Mandelstam-inspired
 model 
of Sec. \ref{LFM}  at $|\xi|=1$ (see text). Right Panel: the same as in the
Left Panel, but for the isovector GPD.}
\label{Hvmfig} 
\end{figure}

The previous figures have illustrated "integral" properties of the pion GPD's, 
like
em form factor and the generalized ones, or 
 the parton distribution, i.e. $H^{I=1}(x,0,0)$. In the following figures, 
 the
isoscalar and isovector GPD's are shown  in the plane $(x,t)$ with
 $-1\leq x\leq 1$ and $-10$ (GeV/c)$^2\leq t \leq 0$, but with  fixed
values for $\xi $, as dictated by the two phenomenological models, namely 
$|\xi|=1$ for the Mandelstam-inspired model (Sec. \ref{LFM}) 
and $\xi=0$ for the
LFHD model (Sec. \ref{LFHD}), respectively.  
The covariant model (Sec. \ref{SYM}), 
in its two versions for the momentum 
dependence (Eqs. (\ref{vertexs}) and (\ref{vertexp})), will be compared 
to the results for the two phenomenological models, that, in some sense,  
represent  two extrema, in the Fock language: the first model is 
 basically related to
the non-valence (ERBL) region, the second one is related to the valence (DGLAP)
 domain.
In order to cover the whole range of $\xi$ for the given interval of $t$ 
(i.e. $-10$ (GeV/c)$^2\leq t \leq 0$)  the covariant model has been
evaluated by assuming $m_\pi=0$, as already pointed out.  Finally, let us 
stress that the GPD's are divided by
$F_{mon}$, as in the case of the em form factor, for avoiding log plot and for
emphasizing as many details as possible.
 In Fig. \ref{Hsymsfig}, the results 
of the covariant symmetric model  are shown for
$|\xi|=1$, in order to be compared with the calculations performed by using the  
Mandelstam-inspired
 model, presented in Fig. \ref{Hvmfig}. We remind that the phenomenological
 model has 
a  photon-quark vertex  dressed by a microscopical VMD, as discussed in Sec.
\ref{LFM}. 
 \begin{figure}[t]
\includegraphics[width=8.0cm]{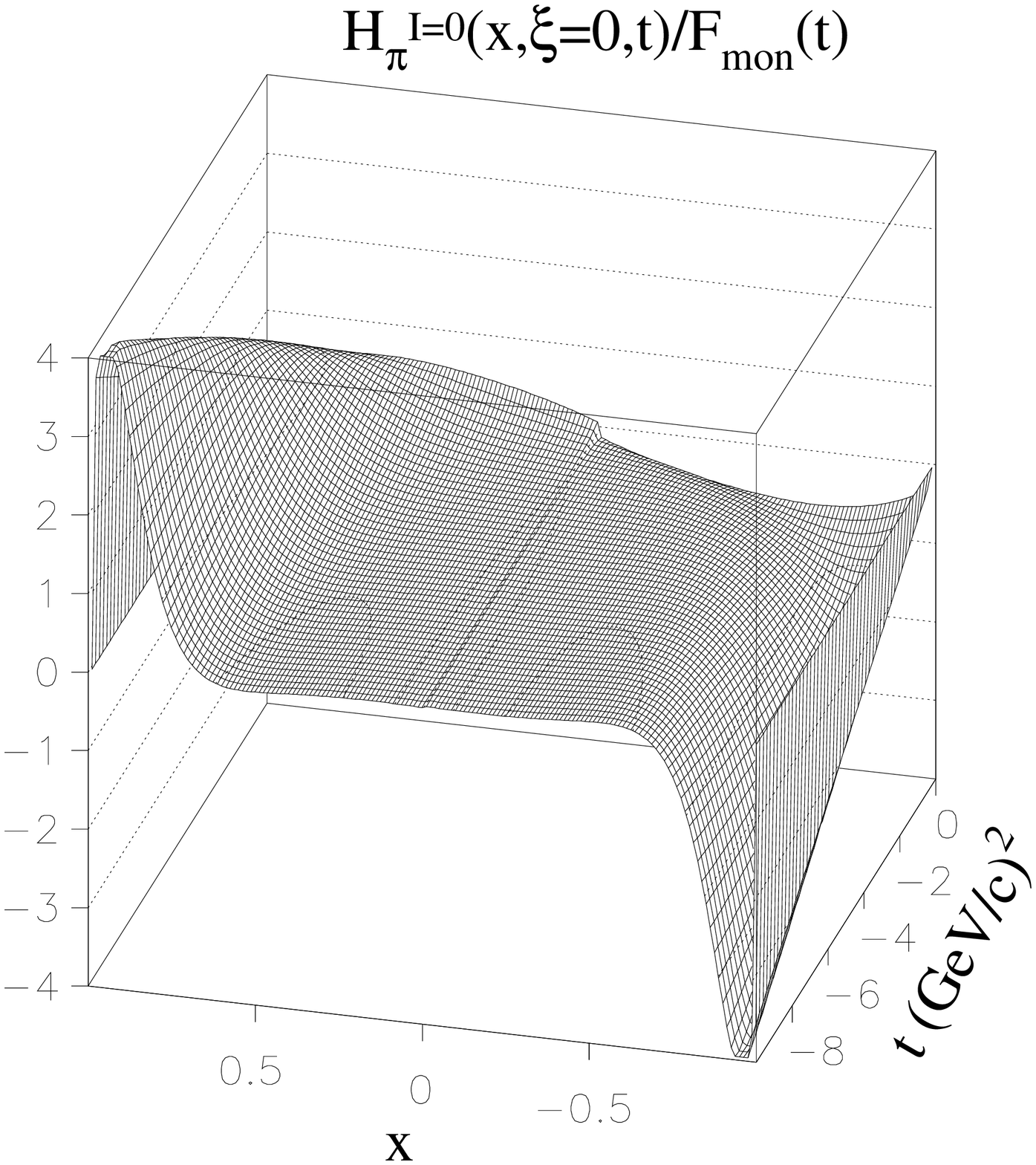}~~
~\includegraphics[width=8.0cm]{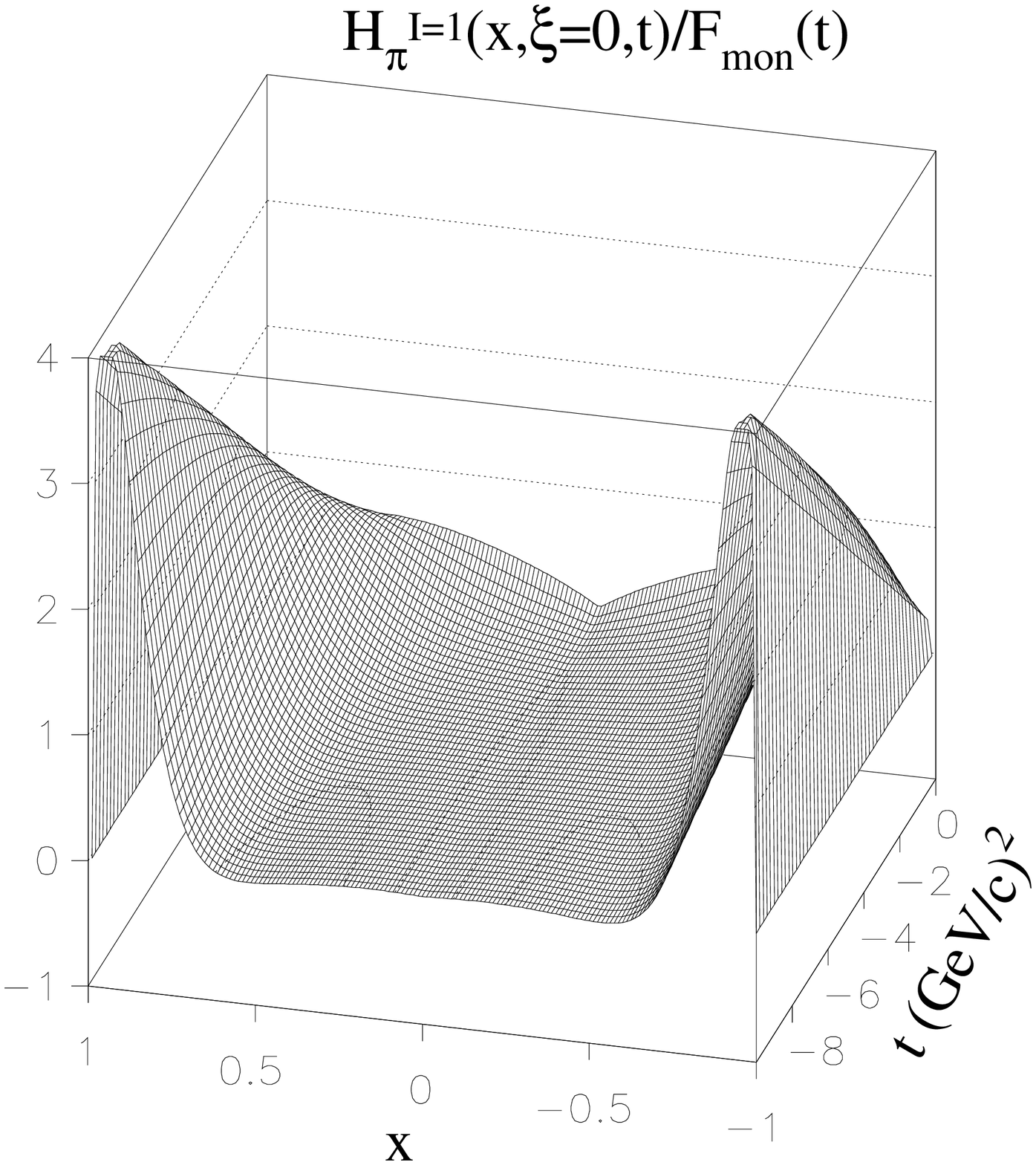}

\vspace{-3cm}\includegraphics[width=8.0cm]{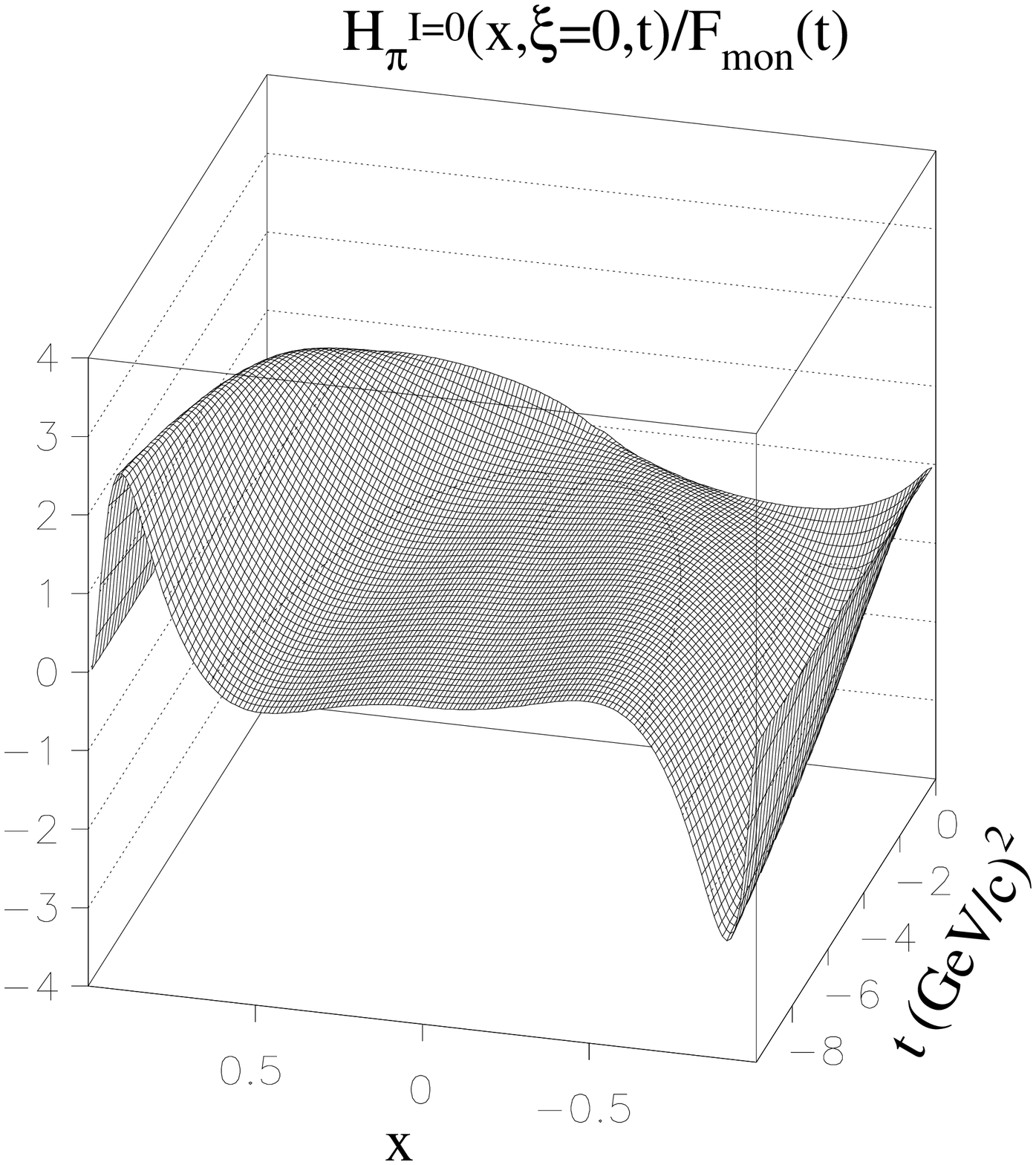}~~~
\includegraphics[width=8.0cm]{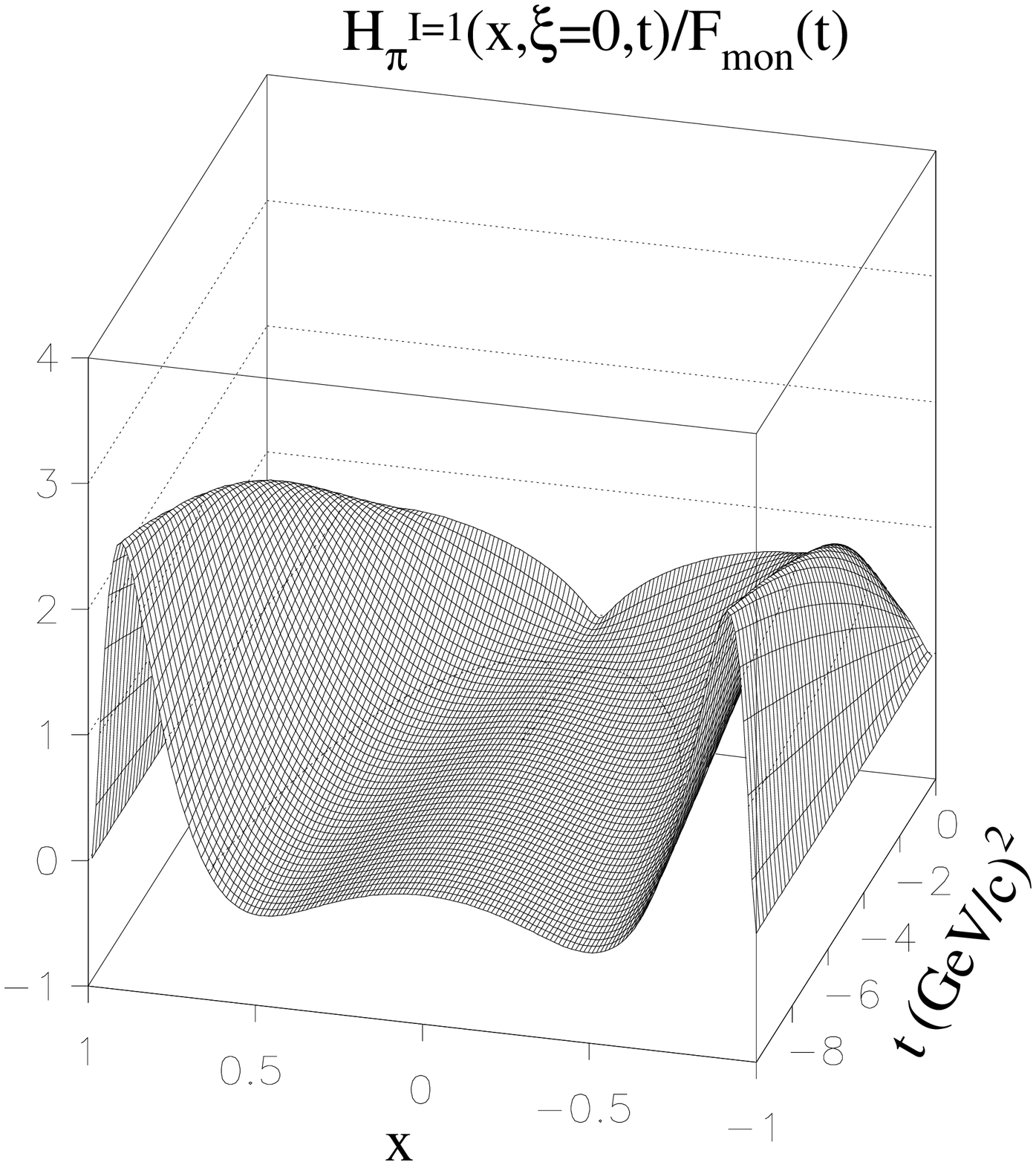}
\caption{Upper Left Panel: Isoscalar no-helicity flip GPD from the covariant 
symmetric model of Sec. \ref{SYM}, with the sum-form for the Bethe-Salpeter
amplitude (Eq. (\ref{vertexs})) at
$\xi=0$, and $m_\pi=0$.  Upper Right Panel: the same as in the
Left Panel, but for the isovector GPD.
Lower Panels: the same as in the Upper Panels, but for the product-form for 
the Bethe-Salpeter
amplitude (Eq. (\ref{vertexp})).}
\label{Hsxi0fig} 
\end{figure}
\begin{figure}
\includegraphics[width=8.0cm]{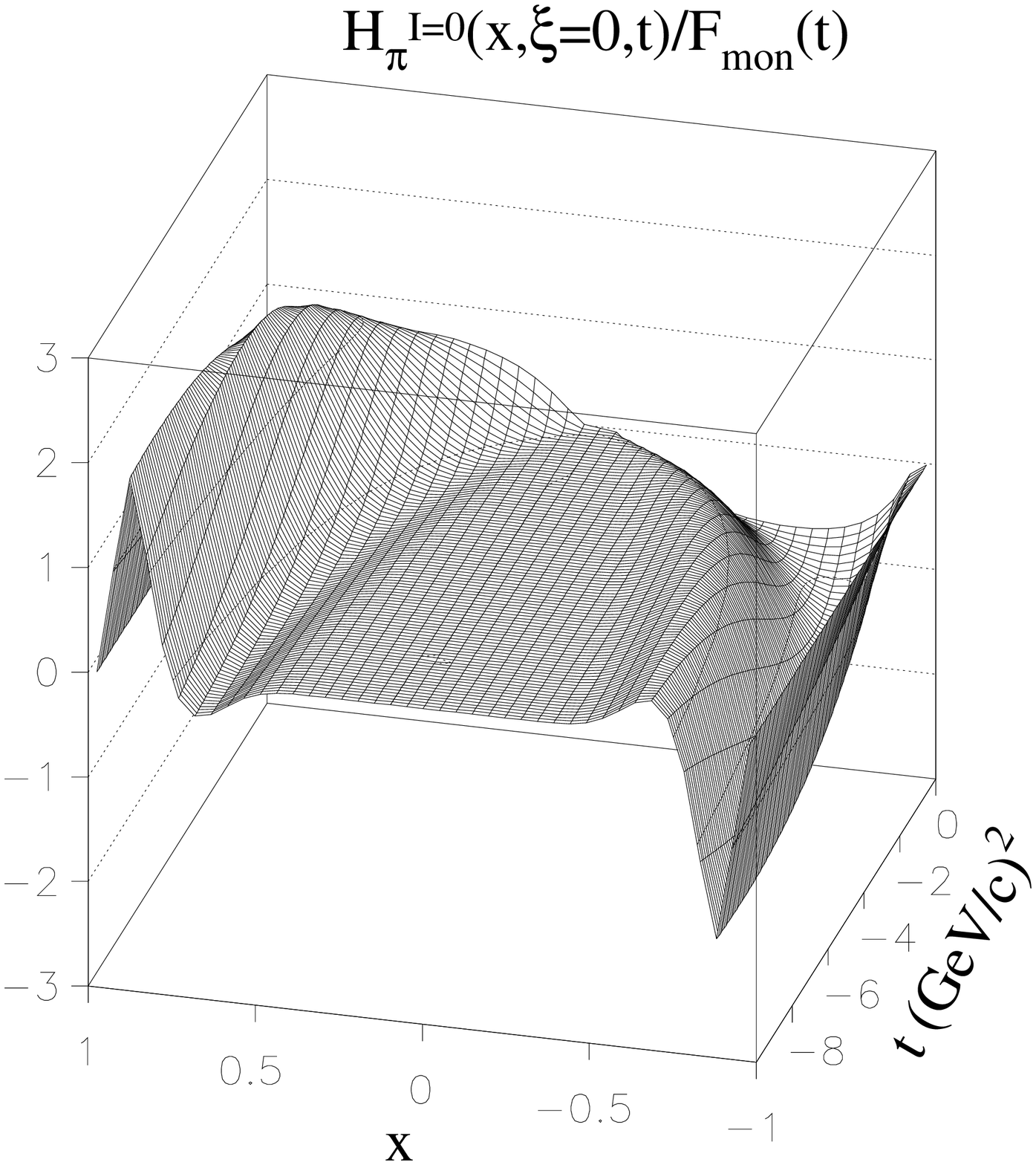}~~~\includegraphics[width=8.0cm]
{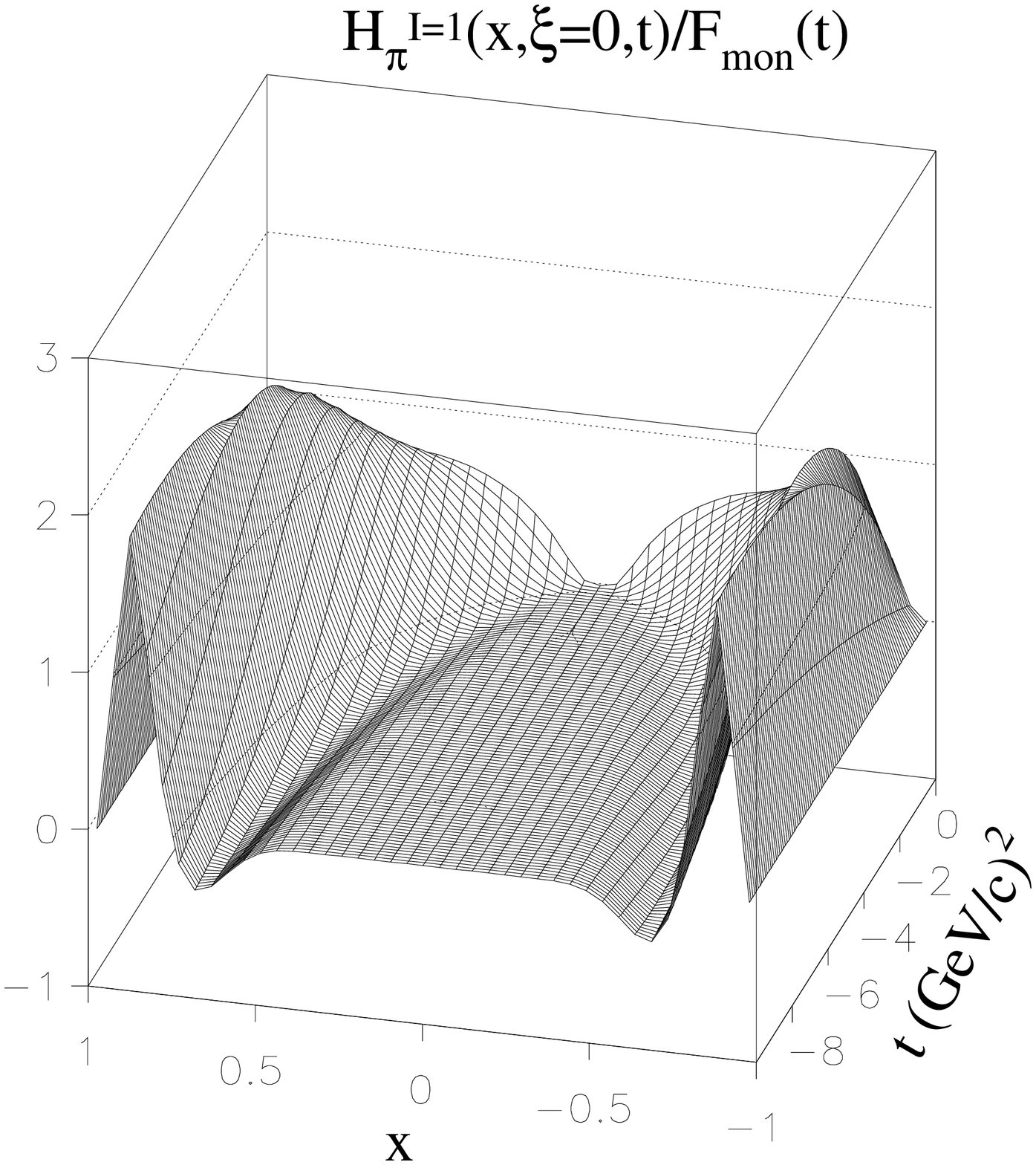}
\caption{Left Panel: Isoscalar no-helicity flip GPD from the  LFHD model 
 of Sec. \ref{LFHD} at $\xi=0$ (see text).
Right Panel: the same as in the
Left Panel, but for the isovector GPD.}
\label{Hbrbfig} 
\end{figure}
\begin{figure}
\includegraphics[width=8.0cm]{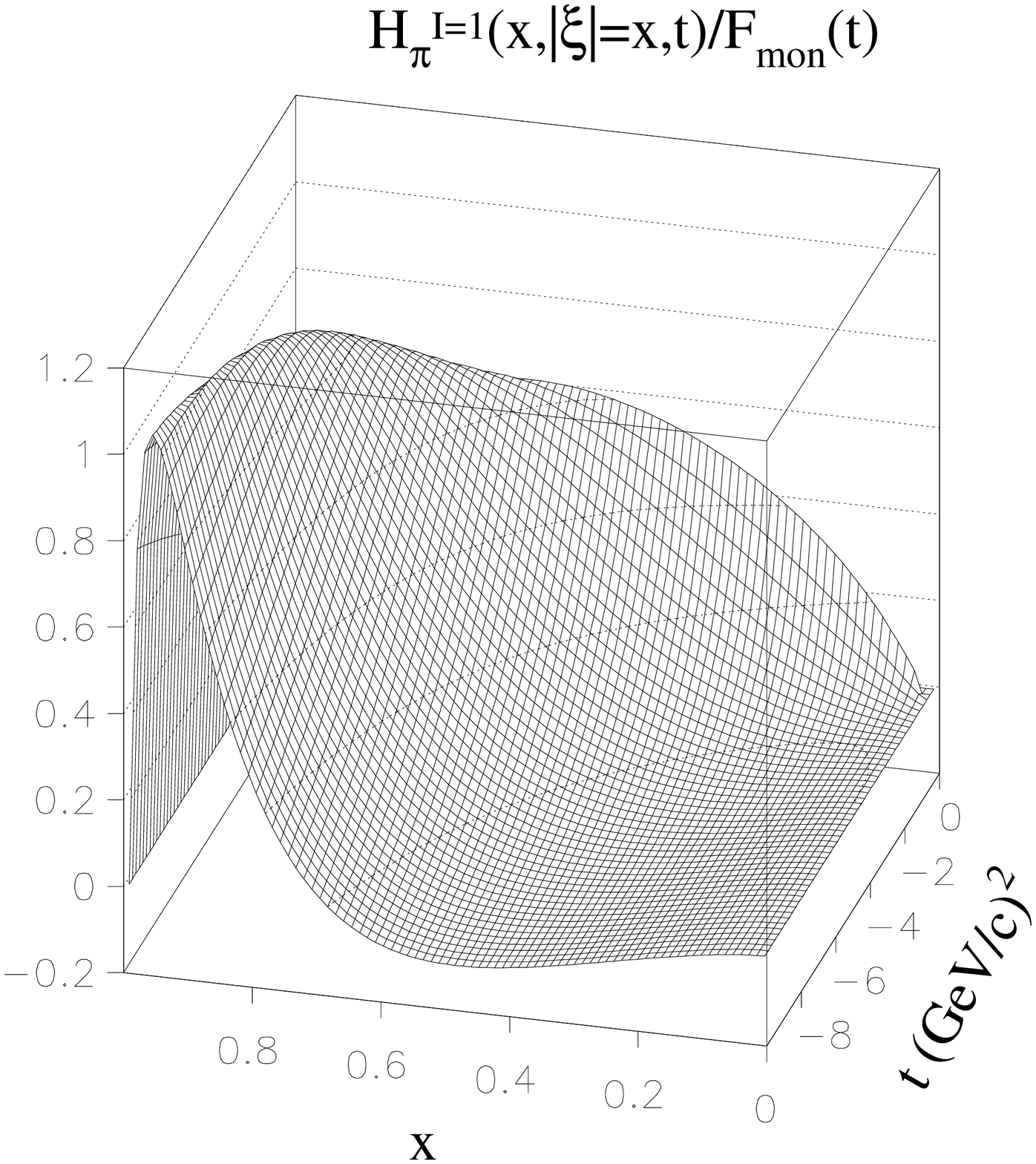}
\caption {Isovector no-helicity flip GPD from the covariant 
symmetric model of Sec. \ref{SYM}, with the product-form for the Bethe-Salpeter
amplitude (Eq. (\ref{vertexp})) at
$|\xi|=x$, and $m_\pi=0$.}
\label{H1xixfig} 
\end{figure}
In Fig. \ref{Hsxi0fig}, the no-helicity flip GPD's
of the covariant symmetric model  are shown for
$\xi=0$, allowing a comparison with the calculations performed by using the  
LFHD
 model, presented in Fig. \ref{Hbrbfig}.  
 For $\xi=0$, where only the valence component is acting and $|x|= x_q$,
  a nice feature, 
 stemming from the figures of  both isoscalar and
 isovector GPD's,  is shared by all the presented models: in the
 limit of large $|t|$ the collinearity clearly emerges, as shown by the
 migration of
 the maximum (minimum) value from $|x|\simeq 1/2$ for $t=0$ toward $|x|\sim 1$ 
 for $|t|\to \infty$. Such a behavior can be easily  understood in
 the LFHD model, since  
 non vanishing contributions to GPD (cf Eq. (\ref{lfhdgpd})) can be obtained if 
 ${\bm \kappa}^\prime_\perp$ in Eq. (\ref{eq:initialconf}) does not depend too much from 
 ${\bm \Delta}_\perp$, namely
 $x\sim 1$ (notice that for $x$  exactly 1, the free mass blows up and the 
 wave functions
 become vanishing, as well as  GPD's).  
  Correspondingly, for $|\xi|=1$,
  where only the non-valence component
  is acting, the relevance of the $x$ region around $\pm 1$ can be explained by
  the pair-production mechanism. For simplicity, let us consider large values of
  $|t|$, that amount to large values of $\Delta^+=\Delta_z$ (remind that in
  the Breit frame $\Delta^0=0$). Then, using $\Delta^+=k^+_q+k^+_{\bar q}=
  k_{z q}+k_{z\bar q}\sim 2 k_{zq}\ge 0$ (given our choice for the sign of
  $\Delta^+$),  and the fact that each quark in the pair is almost on its
  mass-shell,  we can approximate $2k^+=k^+_q-k^+_{\bar q}\sim
  2E_q=2\sqrt{m^2+|\vec k_q|^2}$. Thus, one can see that,  
  when $\xi=-1$, $x$
  becomes close to 1 for $\Delta_z>> m$, since $x=k^+/P^+=
  2k^+/\Delta^+\sim E_q/k_{zq} \to  1$. The case $x=-1$ can be obtained for
  $\Delta^+\le 0$.
  
  Finally in Fig. \ref{H1xixfig}, the isovector GPD, evaluated within the
  covariant model adopting the product-form of Eq. (\ref{vertexp}) and
  $m_\pi=0$, is shown for the case $x=|\xi|$ and $0 \geq t \geq -10 $ (GeV/c)$^2$.
  This kinematical region, where the transition from DGLAP to ERBL regimes
  occurs, should be relevant for the
  experimental studies of the single spin asymmetry (see, e.g. the discussion in
  \cite{diehlpr}). 
 Let us notice that in our covariant analytic model the GPD is continuous
  at $x=|\xi|$.
 
From the 3D plots, one can see that the covariant model, in the version
with  the product-form for the momentum-dependent part of the BS amplitude,
 is able to reproduce quite satisfactorily the GPD's evaluated within the two
 phenomenological models, the Mandelstam-inspired and the LFHD ones, and therefore one could argue that it contains the main
 ingredients for a realistic descriptions of the constituents inside the pion.
 In view of this, it appears challenging to test the covariant model (or its
 refinements \cite{nafpps}  based on the Nakanishi representation, see, e.g.
 \cite{carbonellepja}, for a
 recent applications to a bosonic system) of the BS amplitude, 
 in comparisons with experimental data, whose analysis requires
 the knowledge of the pion GPD's.

\section{Conclusion}
\label{END}
In this paper, we have investigated the no-helicity flip Generalized Parton
Distributions of the pion by using three models, based on a description of the
pion where constituent quarks with masses between  $200$ MeV and $250$ MeV are
considered. In particular, we have evaluated the isoscalar and isovector GPD's 
adopting a covariant, 
analytic model and two Light-front phenomenological models. It is important to notice that 
the first model, based on 4D Ansatzes for the Bethe-Salpeter amplitudes,
 allows us to explore the whole
kinematical domain of the three variables $x$, $\xi$ and $t$ upon which the
GPD's depend, while the others two are, presently, constrained to a given value
of $\xi$. The second model, the Mandelstam-inspired model of Sec. \ref{LFM},
 is a natural extension
of the approach proposed in Ref. \cite{DFPS} for a successful investigation of the em form
 factor of
the pion in both the space- and timelike regions.
 Main features of the model are: i) a microscopical Vector Meson
Model dressing for
 the quark-photon vertex and ii) proper Ansatzes for the 3D LF projection of
the BS amplitudes of both pion and vector mesons, taken as the eigenfunctions of
a LF square mass operator \cite{FPZ}. 
 As in \cite{DFPS}, the  
assumption $m_\pi=0$ is added, and this  simplification allows calculations of the GPD's only for 
the value  $|\xi|=1$ 
(cf Eq.
(\ref{nompi})), namely  the non-valence region covers the whole range $1\geq x\geq
-1$.

On the contrary, the LFHD model of Sec. \ref{LFHD} is based on a Poincar\'e
covariant description of the pion, with a proper treatment of the spin wave
functions, due to the presence of the Melosh rotations. The momentum part
 of the
pion wave function is given by a Gaussian  function, that 
contains the dynamical input of the model through  
two adjusted parameters. A bare quark-photon vertex is assumed.
It is worth noting that  the model yields a
description of the GPD's for $\xi=0$, i.e. the  valence region can be
investigated. 

The covariant symmetric model of Sec. \ref{SYM}, based on a Mandelstam formula
for matrix elements of the operators yielding the isoscalar and isovector GPD's,
allows us to have close expressions for  the physical quantities, since analytic
forms for the momentum-dependent part of the Bethe-Salpeter amplitude are
adopted and a bare quark-photon vertex is assumed as well. Such a covariant
model can be applied  for any value of $x$, $\xi$ and  $t$, and therefore can be
used for interpolating between the  two phenomenological models. A peculiar
 feature is given by the presence of  instantaneous terms, both in
the valence and non-valence regions, since we fully take into account the
analytic structure of the BS amplitude.

The comparison with the em form factor (Fig. \ref{ffpfig}) suggests that 
the covariant model with a
sum-form of the BS amplitude has a non realistic increasing behaviour with respect to
$F_{mon}$, for large $|t|$, which    leads to a divergent density at short
distances,
while the version with a product-form together with the LFHD model decrease 
more
rapidly than $F_{mon}$. Finally, the Mandelstam-inspired model and the Lattice results
(red curve in Fig. \ref{ffpfig}),
arbitrarily extended from $t=-4$ (GeV/c)$^2$ to $t=-10$ (GeV/c)$^2$, given the
analytic form proposed in \cite{brom07} for extrapolating the Lattice data to
the physical $m_\pi$, show a moderate decreasing with respect to  $F_{mon}$, for 
 large $|t|$.
 Such a comparison  for the em form factor and the analysis of the parton distribution in Fig.
\ref{strucx}
point to the relevance of the behavior of the pion valence function (or 
better the momentum part of the BS amplitude) for large
transverse momentum. In particular the product-form, that has a behavior
at large transverse momentum $|{\bf k}_\perp|$ compatible with the one suggested
by the one-gluon-exchange dominance (see, e.g. \cite{carbonellepja}), seems to give a
consistent description of both the tail of the em form factor and the end-point
fall-off of the parton distribution. With respect to this finding, more details
can be gained  from the
investigation of the chiral-even  transverse-momentum dependent distribution, as shown in
Fig. \ref{f1xkfig}.

 Another important step in the characterization of the covariant
model is given by the comparison of the generalized form factors with the
Lattice results. For the present, the comparison is restricted to the
gravitational form factors,
$A^{I=0}_{2,0}(t)$ and $A^{I=0}_{2,2}(t)$, that appear in the second moment of the 
isovector  GPD, $H^{I=0}$, (cf
Eqs. (\ref{gravff}) and (\ref{oddm})).  Indeed, for $A^{I=0}_{2,0}(t)$
 we have
presented results from both our covariant model and  the LFHD approach, while for
$A^{I=0}_{2,2}(t)$  only the covariant calculations are available (let us remind that
calculations with $\xi \ne 0$ are necessary for disentangling 
both form factors). Unfortunately, since    Lattice data have been 
obtained in a 
$t$-interval   not too wide and   are affected by large
 uncertainties, one cannot yet draw   stringent conclusions from the comparison
 shown in Fig. \ref{figsecm}. However,
 the   encouraging agreement between model calculations and Lattice
 data for both   ratios, $A^{I=0}_{0,2}(t)/A^{I=0}_{0,2}(0)$ 
 and $A^{I=0}_{2,2}(t)/A^{I=0}_{2,2}(0)$, suggests to extend our analysis also 
 to the spin-flip
GPD's, since Lattice results are available for the lowest moments
 \cite{brommelprl}, in order to  explore the onset of the dominance of a
 one-gluon-exchange mechanism  for a light
 hadron. 

To complete our analysis, we have studied the GPD's in the $(x,t)$ plane
for fixed values of $\xi$, i.e. $|\xi|=0,1,x$.  
These values  are representative of different, interesting cases. 
The first one, $\xi=0$, involves
contributions to GPD's only in the valence region, while the second one 
involves
 contributions only from the non-valence one. Finally, the case $|\xi|=x$ illustrates the
transition from the DGLAP region to the ERBL one.
The covariant model can explore the whole 3D
space of the variables $(x,\xi,t)$ and it is compared with LFHD model for
$\xi=0$, and with the Mandelstam-inspired model for $|\xi|=1$, while for $|\xi|=x$ shows a
smooth transition from DGLAP region to the ERBL one, given the continuity of the model.
It should be pointed out that the covariant model with the product-form for the
Bethe-Salpeter amplitude exhibits  an overall agreement with
 the Mandelstam-inspired model, for $|\xi|=1$, 
and with
the LFHD model, for $\xi=0$. Therefore, from these findings 
one could conjecture that the general shape, illustrated by the previous covariant
model and  the phenomenological
 ones, is a typical feature of the pion GPD's,
dictated from both kinematical arguments (cf the discussion at the end of Sec. 
\ref{Ris}) and the dynamical input reflected by the proper fall-off of   the momentum
distribution (cf the one-gluon exchange dominance at short distances).

Further analyses, to make more and more realistic the models presented in this
paper, are in progress.
 \section*{Acknowledgments}
This work was partially supported by the Brazilian agencies CNPq and
FAPESP and by Ministero della Ricerca Scientifica e Tecnologica. 
It is also part of the Research Infrastructure Integrating Activity
``Study of Strongly Interacting Matter'' (acronym HadronPhysics2, Grant
Agreement n. 227431) under the Seventh Framework Programme of the European
Community. 
T. F. acknowledges the hospitality of the Dipartimento di Fisica,
Universit\`a di Roma "Tor Vergata" and of Istituto Nazionale di
Fisica Nucleare, Sezione Tor Vergata and
 Sezione di Roma.

\newpage
 \appendix

\section{Kinematics}
\label{kinea}
Following the notations of Fig. \ref{fig1}, where $\Delta^+\ge 0$, 
one obtains from Eq. (\ref{kin}) that $0 \ge \xi$. 

In the valence region,
 for a quark,  
 one has: i) in the initial state,
$p^+\ge k^+ -\Delta^+/2\ge 0$, i.e.
 $P^+ \ge k^+ \ge \Delta^+/2$, (notice that necessarily the spectator 
 constituent
 is an antiquark, since $0 \ge k^+-P^+$) and  then $1 \ge x\ge -\xi$; ii) 
 in the final state,    $p^{\prime +}\ge k^+ +\Delta^+/2\ge 0$, i.e.
$P^+ \ge k^+ \ge - \Delta^+/2$, and then
$1 \ge x\ge \xi$. 
Therefore 
 in the valence region, one gets  the 
interval $1 \ge x\ge -\xi$, and given our choice for $\xi$ one has 
$1 \ge x\ge |\xi|$.  

For an  antiquark in the initial pion,
 the four-momentum is $k+\Delta/2$, while  the spectator quark has 
four-momentum   $k+P$. In the final pion, the antiquark 
four-momentum is $k-\Delta/2$. The antiquark  plus components are
  negative both in the initial and in the final pion. Therefore i) 
$p^+ \ge -(k^+ +\Delta^+/2) \ge 0$ that leads to $ \xi \ge x \ge -1$  and  ii)  $ p^{\prime +}\ge
-(k^+-\Delta^+/2) \ge 0$, i.e. $ -\xi \ge x \ge -1$. 
Summarizing, for an 
antiquark
in the valence region one finds $ -|\xi| \ge x \ge -1$.

In the non-valence region, one has to deal with  a $q\bar q$  production, i.e.
$0>k^+ - \Delta^+/2$ and $ k^+ + \Delta^+/2>0$ (see Fig. \ref{figbs}),  and those constraints
translate into $\xi < x < -\xi$. 
The   $q\bar q$ annihilation  is prevented by the  choice
of a positive $\Delta ^+$. 
In order to have general extrema, holding for both
positive and negative $\Delta^+$, one can write $|\xi|> x > -|\xi|$.

\section{ Integration on $k^-$}
\label{intkm}

In this Appendix,  the  no-helicity flip  GPD  for the symmetric covariant 
models (see Sec. \ref{SYM}) calculated using 
Eq. (\ref{jmu})
 and  the momentum dependent part of the BS amplitude, given by
 Eqs. (\ref{vertexs}) or   (\ref{vertexp}).

The evaluation of the trace in Eq. (\ref{trace})  can be simplified
according to the decomposition of the Dirac propagator shown in Eq.
(\ref{inst}) and reminding that $[\gamma^+]^2=0$. By introducing the variable 
$\kappa=P-k$, 
one has 
 \be
Tr[{\cal O}^+(\kappa^-)]=Tr\left \{ \left(\psla{\kappa}+m\right)
 \left(\psla p^\prime-\psla{\kappa}+m\right) 
\gamma^+\left(\psla p-\psla{\kappa}+m\right)\right \}=
\nonu =
Tr[{\cal O}^+(\kappa^-_{on})]+
{\left (\kappa^- - \kappa^-_{on}\right)\over 2}~ Tr\left \{ \gamma^+
\left[ \left(\psla p^\prime-\psla{\kappa}\right )_{on}+m\right]~ 
\gamma^+~\left[\left(\psla p-\psla{\kappa}\right )_{on}+m\right]\right \}=
\nonu=-4~\left\{ \kappa^+~\left [(p^\prime-\kappa)_{on} 
\cdot (p-\kappa)_{on}-m^2 \right ]
-(p^{\prime +}-\kappa^+) ~\left [\kappa_{on} \cdot 
(p-\kappa)_{on}-m^2 \right ] 
+ \right. \nonu \left . -(p^+-\kappa^+)~
 \left [(p^\prime -\kappa)_{on} \cdot \kappa_{on}-m^2 \right] \right \}+
4~\left (\kappa^- - \kappa^-_{on}\right)~  
\left(  p^{\prime +}-\kappa^+ \right )~\left(  p^+ -\kappa^+ \right )
\ee
where
\be
Tr[{\cal O}^+(\kappa^-_{on})]=Tr\left \{ \left(\psla{\kappa}_{on}+m\right)
\left[ \left(\psla p^\prime-\psla{\kappa}\right )_{on}+m\right]~ 
\gamma^+~\left[\left(\psla p-\psla{\kappa}\right )_{on}+m\right]\right \}
\ee
After performing the scalar products, one gets 
\be
Tr[{\cal O}^+(\kappa^-)]=4~p^{\prime +}p^{ +}~\kappa^-_{on}
-
{\kappa^+~|{\bm \Delta}_\perp|^2}+
2 \Delta^+~{\bm \kappa}_\perp \cdot 
 {\bm \Delta}_\perp + \nonu +4~\left (\kappa^- - \kappa^-_{on}\right)~  
\left(  p^{\prime +} -\kappa^+ \right )~\left(  p^+ -\kappa^+ \right )
\ee

 Given the simple expression adopted for the
momentum dependence of the BS amplitude (see Eqs. (\ref{vertexs}) and
 (\ref{vertexp})),
the analytic integration on $k^-$ can be easily performed in Eq. (\ref{jmu}).

By using the LF variables (i.e. $d^4 \kappa \to d\kappa^+d\kappa^- d{\bm
\kappa}_\perp/2$) one obtains
\be
 H^u(x,\xi,t) = -\imath N_c~{\cal R}~\int
{d\kappa^+d\kappa^- d{\bm
\kappa}_\perp \over 4(2\pi)^4}~\delta\left[P^+(1-x)-\kappa^+\right] 
{Tr[{\cal O}^+(\kappa^-)] \over \kappa^+~(p^+-\kappa^+)~(p^{\prime +}-\kappa^+)}
~\times \nonu
{1 \over \left(\kappa^- -\kappa^-_{on}+i{\epsilon\over \kappa^+}\right)}
{1 \over \left [p^- -\kappa^- -(p-\kappa)^-_{on}+i { \epsilon \over
 (p^+ - \kappa^+)} \right ]}
 {1 \over \left [p^{\prime -} -\kappa^- -(p^\prime-\kappa)^-_{on}+i
 {\epsilon\over
 (p^{\prime +}-\kappa^+)}\right ]}\times \nonu
 \Lambda(\kappa,p^{\prime})~
\Lambda(\kappa,p) \label{a1}
\ee
where
\be
\kappa^-_{on}=\frac{m^2+|\bm \kappa_\perp|^2}{\kappa^+}\quad \quad
(p -\kappa)^-_{on}=\frac{m^2+|\bm p _\perp-\bm \kappa_\perp|^{2}}
{(p^  +-\kappa^+)}\quad \quad 
(p^\prime -\kappa)^-_{on}=\frac{m^2+|\bm p^\prime_\perp-\bm \kappa_\perp|^{2}}
{(p^{\prime +}-\kappa^+)}\nonu
\label{pol1}
\ee
 In the integration 
over the minus component $\kappa^-$ one faces with the following six poles 
(coming from the BS
amplitudes and the Dirac propagators)
\be \kappa^-_{1(2)}=\kappa^-_{on(R)}-i{\epsilon\over \kappa^+} \nonu
\kappa^-_{3(4)}=p^- -(p-\kappa)^-_{on(R)}+i{\epsilon\over (p^+-\kappa^+)}
 \nonu
\kappa^-_{5(6)}=p^{\prime -} -(p^\prime-\kappa)^-_{on(R)}+
i{\epsilon\over (p^{\prime +}-\kappa^+)}
 \ee
where $\kappa^-_R$, $(p-\kappa)^-_{R}$ and $(p^\prime -\kappa)^-_{R}$ can be
obtained from the corresponding quantities in Eq. (\ref{pol1}) by substituting
$m \to m_R$. Notice that  $\kappa^-_{2}$ can appear both as a single and as a 
double
pole.

It is easily seen that the analytic integral  (\ref{a1}) is not vanishing only
if $p^{\prime +}\ge \kappa^+\ge 0$. Furthermore we can recognize two subinterval
i) $p^{ +}\ge \kappa^+\ge 0$, or valence region, and ii) $p^{\prime +}\ge
\kappa^+\ge p^{ +}$, the non-valence region.
Let us stress that Eq. (\ref{a1}) is vanishing for $x < ~-|\xi|$, since in
this case $\kappa^+ =P^+~(1-x)> P^+~(1+|\xi|)=p^{\prime +}$.

In the valence region, only the poles $\kappa^-_{1}$ and $\kappa^-_{2}$
 belong to
the lower semiplane. In the non-valence region, only $\kappa^-_{5}$ and 
$\kappa^-_{6 }$ belong to
the upper semiplane.

To obtain the no-helicity flip GPD in the valence region, let us integrate over  
$\kappa^-$ closing the contour in the lower semiplane. 
The 
contribution from $\kappa^-_{1}$ reads as follows
\be
H_{(v)on}^u(x,\xi,t)=
  -  {N_c~{\cal R}\over 4(2\pi)^3}~
\int d \bm \kappa_{\perp} \int^{p^+}_0 d \kappa^{+} 
{\delta\left[P^+(1-x)-\kappa^+\right] \over \kappa^+(p^+-\kappa^+)
 (p^{^{\prime}+}-\kappa^+)} 
Tr[{\cal O}^+(\kappa^-_{on})]
\nonu \times  
\frac{\left. \Lambda(\kappa,p)\right|_{\kappa^-_{on}}}
{ \left[p^- - \kappa^-_{on} -(p-\kappa)^-_{on}\right]}~
\frac{\left. \Lambda(\kappa,p^\prime)\right|_{\kappa^-_{on}}}
{\left[p^{\prime -} - \kappa^-_{on} -(p^\prime-\kappa)^-_{on}\right]}
\label{hon}\ee
where for the sum-form, Eq. (\ref{vertexs}), one has
\be
\left. \Lambda(\kappa,p)\right|_{\kappa_{on}}= C_1~\left \{
\frac{1}
{(p^+-\kappa^+)\left[p^- - \kappa^-_{on} -(p-\kappa)^-_R\right]}+
\frac{1}
{\kappa^+(\kappa^-_{on} - \kappa^-_{R} )}\right \}
\label{a13}
\ee
and for the product-form, Eq. (\ref{vertexp}), one has
\be
\left. \Lambda(\kappa,p)\right|_{\kappa_{on}}= C_2~
\frac{1}
{(p^+-\kappa^+)\left[p^- - \kappa^-_{on} -(p-\kappa)^-_R\right]}~
\frac{1}
{\kappa^+(\kappa^-_{on} - \kappa^-_{R} )}
\label{a13p}
\ee
For 
the sum-form, the pole $\kappa^-_{R}$ generates  a contribution 
as a single pole and a  contribution as a double 
pole. 

The single-pole contribution is given by 
\be
H_{(v)1}^u(x,\xi,t)=- {N_c~{\cal R}\over 4(2\pi)^3}~C_1^2
\int d \bm \kappa_{\perp} \int^{p^+}_0 d \kappa^{+} 
{\delta\left[P^+(1-x)-\kappa^+\right] \over (\kappa^+)^2(p^+-\kappa^+)
 (p^{^{\prime}+}-\kappa^+)} 
Tr[{\cal O}^+(\kappa^-_{R})]~
 \times \nonu {1 \over (\kappa^-_{R} - \kappa^-_{on})}~
\frac{1}
{\left [ p^- - \kappa^-_{R} -(p-\kappa)^-_{on})\right ]~
\left [p^{\prime -} - \kappa^-_{R} -(p^\prime-\kappa)^-_{on}\right ]}~
\times \nonu\left \{\frac{1}
{(p^{+}-\kappa^+)\left [ p^- - \kappa^-_{R} -(p-\kappa)^-_{R}\right ]}+
\frac{1}
{(p^{^{\prime}+}-\kappa^+)\left [ p^{\prime -} - \kappa^-_{R} -
(p^\prime-\kappa)^-_{R}\right ]}
\right \} \label{i1}\ee
and the  double-pole contribution is given by
\be
H_{(v)2}^u(x,\xi,t)=- {N_c~{\cal R}\over 4(2\pi)^3}~C_1^2
\int d \bm \kappa_{\perp} \int^{p^+}_0 d \kappa^{+} 
{\delta\left[P^+(1-x)-\kappa^+\right] \over (\kappa^+)^3(p^+-\kappa^+)
 (p^{^{\prime}+}-\kappa^+)}~\times \nonu 
{d \over d \kappa^-}\left. \left \{ {Tr[{\cal O}^+(\kappa^-)]
\over (\kappa^- - \kappa^-_{on})}~{1\over 
\left [ p^- - \kappa^- -(p-\kappa)^-_{on})\right ]~\left [
p^{\prime -} - \kappa^- -(p^\prime-\kappa)^-_{on}\right ]}\right \}
\right|_{\kappa^-_{R}}\label{i2}\ee

For the product-form, the pole $\kappa^-_{R}$ generates only a double-pole 
contribution, given by  
\be
H_{(v)2^\prime}^u(x,\xi,t)=- {N_c~{\cal R}\over 4(2\pi)^3}~C_2^2
\int d \bm \kappa_{\perp} \int^{p^+}_0 d \kappa^{+} 
{\delta\left[P^+(1-x)-\kappa^+\right] \over (\kappa^+)^3(p^+-\kappa^+)^2
 (p^{^{\prime}+}-\kappa^+)^2}~\times \nonu 
{d \over d \kappa^-}\left. \left \{ {Tr[{\cal O}^+(\kappa^-)]
\over (\kappa^- - \kappa^-_{on})}~\frac{1}
{\left [ p^- - \kappa^- -(p-\kappa)^-_{on})\right ]~\left [
p^{\prime -} - \kappa^- -(p^\prime-\kappa)^-_{on}\right ]}\times \right . \right .
\nonu \left . \left .
\frac{1}
{\left [ p^- - \kappa^- -(p-\kappa)^-_{R})\right ]~\left [
p^{\prime -} - \kappa^- -(p^\prime-\kappa)^-_{R}\right ]}\right \}
\right|_{\kappa^-_{R}}\label{i3}\ee
The  contribution in the non-valence region can be evaluated by considering the poles 
$\kappa^-_{5}$ and 
$\kappa^-_{6 }$. In particular the contribution from  $\kappa^-_{5}$
has the same form for both choices of the BS amplitudes, i.e.
\be
H_{(nv)5}^u(x,\xi,t)=
  - {N_c~{\cal R}\over 4(2\pi)^3}~
\int d \bm \kappa_{\perp} \int^{p^{\prime +}}_{p^{ +}} d \kappa^{+}
{\delta\left[P^+(1-x)-\kappa^+\right]\over \kappa^+(p^+-\kappa^+) 
(p^{^{\prime}+}-\kappa^+)}~\times \nonu 
 \frac{Tr[{\cal O}^+(p^{\prime-}
-(p^\prime-\kappa)^-_{on})]~\left. \Lambda(\kappa,p^\prime)\right|_{p^{\prime-}
 -(p^\prime-\kappa)^-_{on}}~\left. \Lambda(\kappa,p)\right|_{p^{\prime-} -(p^\prime-\kappa)^-_{on}}}
{\left [p^{\prime-} -(p^\prime-\kappa)^-_{on}-\kappa^-_{on}\right ]
~\left [p^{-} - p^{\prime -}+(p^\prime-\kappa)^-_{on} -(p-\kappa)^-_{on} 
\right ]}
\label{b3}
\ee
while the contributions from $\kappa^-_{6}$, reads differently for the
sum-form, viz
\be
H_{(nv)6}^u(x,\xi,t)=
  -  {N_c~{\cal R}\over 4(2\pi)^3}~C_1~
\int d \bm \kappa_{\perp} \int^{p^{\prime +}}_{p^{ +}} 
d \kappa^{+}{\delta\left[P^+(1-x)-\kappa^+\right]\over \kappa^+(p^+-\kappa^+) 
(p^{^{\prime}+}-\kappa^+)^2}~\times 
\nonu {1 \over \left
[p^\prime-\kappa)^-_{R}-(p^\prime-\kappa)^-_{on}\right ] }~
\frac{Tr[{\cal O}^+(p^{\prime -} -(p^\prime-\kappa)^-_{R})]}
{\left [(p^{\prime -} -(p^\prime-\kappa)^-_{R}- \kappa^-_{on}\right ]
}~ \times \nonu \frac{\left. \Lambda(\kappa,p)\right|_{p^{\prime-} -(p^\prime-\kappa)^-_{R}}}
{\left [p^{-} - p^{\prime -}+(p^\prime-\kappa)^-_{R} -(p-\kappa)^-_{on} 
\right ]}
\label{b5}
\ee
and for  the product-form, viz
\be
H_{(nv)6^\prime}^u(x,\xi,t)=
  -  {N_c~{\cal R}\over 4(2\pi)^3}~C_2~
\int d \bm \kappa_{\perp} \int^{p^{\prime +}}_{p^{ +}} 
d \kappa^{+}{\delta\left[P^+(1-x)-\kappa^+\right]\over (\kappa^+)^2~(p^+-\kappa^+) 
(p^{^{\prime}+}-\kappa^+)^2}
\nonu \times Tr[{\cal O}^+(p^{\prime -} -(p^\prime-\kappa)^-_{R})]
~\frac{1}
{\left [(p^{\prime -} -(p^\prime-\kappa)^-_{R}- \kappa^-_{on}\right ]~
\left [(p^\prime-\kappa)^-_{R}-(p^\prime-\kappa)^-_{on}\right ]
}\nonu \times  \frac{1}{\left [p^{-} - p^{\prime -}+(p^\prime-\kappa)^-_{R} -(p-\kappa)^-_{on} 
\right ]}~  
{\left. \Lambda(\kappa,p)\right|_{p^{\prime-} -(p^\prime-\kappa)^-_{R}}
\over \left [(p^{\prime -} -(p^\prime-\kappa)^-_{R}- \kappa^-_{R}\right ]}
\label{b5p}
\ee
Summarizing, for the sum-form one has
\be
 H^u(x,\xi,t)=
 \theta(x-|\xi|)~\theta(1-x)~\left[ H_{(v)on}^u(x,\xi,t)+H_{(v)1}^u(x,\xi,t)+H_{(v)2}^u(x,\xi,t)\right] 
 +\nonu +\theta(|\xi|-x)~\theta(|\xi|+x)\left[H_{(nv)5}^u(x,\xi,t)+H_{(nv)6}^u(x,\xi,t)\right]
\ee
with $H_{(nv)6}^u$ given by Eq. (\ref{b5}), while for the product-form one gets
\be
 H^u(x,\xi,t)=
 \theta(x-|\xi|)~\theta(1-x)~\left[ H_{(v)on}^u(x,\xi,t)+H_{(v)2^\prime}^u(x,\xi,t)\right] 
 +\nonu +\theta(|\xi|-x)~\theta(|\xi|+x)\left[H_{(nv)5}^u(x,\xi,t)+H_{(nv)6^\prime}^u(x,\xi,t)\right]
\ee
with $H_{(nv)6^\prime}^u$ given by Eq. (\ref{b5p}).
\section{Electromagnetic form factor }
\label{ffpiap}
The pion electromagnetic form factor is defined by
\be
F_\pi(t)= {1 \over 2 ~P^+} ~\langle \pi^+(p^\prime)|J(0)\cdot n|\pi^+(p)\rangle
=
\int_{-1}^1  dx\, {H}^{I=1}(x,\xi,t) =\nonu ={1 \over 2}\int_{-\infty}^\infty  dx 
\int \frac{dz^-}{2\pi} e^{i x P^+ z^-}
\times
 \left . \langle \pi^+ (p') | \bar \psi_q (-\frac12 z)
\gamma \cdot n \,\,\tau_3  \psi_q (\frac12 z) | \pi^+ (p) \rangle
\right |_{\tilde z=0}\ee
where the   range of $x$ has been extended from $[-1,1]$ to $[-\infty,\infty]$,
since $H^{I=1}(x,\xi,t)$ is vanishing outside the support $[-1,1]$,  given 
the presence of the delta-function in Eq. (\ref{jmu}) and the kinematical 
relations in
Eq.  (\ref{kin}) (see, e.g.
\cite{diehlpr,Ji98}).

\end{document}